\documentclass[sigplan,screen,10pt]{acmart}

\pdfoutput=1

\AtBeginDocument{%
  \providecommand\BibTeX{{%
    \normalfont B\kern-0.5em{\scshape i\kern-0.25em b}\kern-0.8em\TeX}}}

\setcopyright{acmcopyright}
\copyrightyear{2020}
\acmYear{2020}

\acmConference[Onward! '20]{Proceedings of the 2020 ACMSIGPLAN International
Symposium on New Ideas, New Paradigms,and Reflections on Programming and
Software (Onward! '20)}{November 15--20, 2020}{Chicago, IL}

\usepackage{amsfonts}
\usepackage{amsmath}
\usepackage{mathtools}
\usepackage{prettyref}
\usepackage{listings}
\usepackage{tikz}
\usepackage{subcaption}
\usepackage{ifthen}

\newcommand{\pref}{\prettyref}
\newrefformat{thm}{Theorem~\ref{#1}}
\newrefformat{cor}{Corollary~\ref{#1}}
\newrefformat{lem}{Lemma~\ref{#1}}
\newrefformat{cha}{Chapter~\ref{#1}}
\newrefformat{sec}{Section~\ref{#1}}
\newrefformat{app}{Appendix~\ref{#1}}
\newrefformat{tab}{Table~\ref{#1}}
\newrefformat{fig}{Figure~\ref{#1}}
\newrefformat{alg}{Algorithm~\ref{#1}}
\newrefformat{exa}{Example~\ref{#1}}
\newrefformat{def}{Definition~\ref{#1}}
\newrefformat{li}{Line~\ref{#1}}
\newrefformat{eq}{Equation~\ref{#1}}

\usepackage{xcolor}

\usetikzlibrary{fit,calc}

\colorlet{pink}{red!40}

\colorlet{cyan}{cyan!60}

\newcommand\TC{\textsc{Sifter}}

\newcommand*{\stringobjblue}[1]{{\color{blue}{\texttt{#1}}}}
\newcommand*{\stringobjred}[1]{{\color{red}{\texttt{#1}}}}
\newcommand*{\stringobj}[1]{\ifthenelse{\equal{#1}{?}}{\stringobjred{#1}}{\stringobjblue{#1}}}
\newcommand*{\stringpair}[2]{\stringobj{#1}$\to$\stringobj{#2}}
\DeclareMathOperator*{\NextTo}{NextTo}
\newcommand\NextToLeft{\textrm{NextTo:Left}}
\newcommand\NextToRight{\textrm{NextTo:Right}}
\newcommand\ROne{\textrm{R:1}}
\newcommand\RTwo{\textrm{R:2}}
\newcommand\Rn{\textrm{R:n}}
\newcommand\Ri{\textrm{R:i}}
\DeclareMathOperator*{\LetterSame}{LetterSame}
\DeclareMathOperator*{\LetterSuccessor}{LetterSuccessor}
\DeclareMathOperator*{\LetterPredecessor}{LetterPredecessor}
\DeclareMathOperator*{\IsA}{IsLetterA}
\DeclareMathOperator*{\IsB}{IsLetterB}
\DeclareMathOperator*{\IsNegativeSentiment}{IsNegativeSentiment}
\newcommand{\PlatonicLetter}[1]{Letter:#1}
\newcommand*{\labeltext}[1]{{\color{red}{\texttt{#1}}}}

\newcommand{\scedge}[5]{
    \path[->,ultra thick,draw=#2] (#3) edge
    node[#5,fill=white,fill opacity=0.5,text opacity=1] {$#1$} (#4);
}
\newcommand{\sclr}[4]{
    \scedge{#1}{#2}{#3}{Left}{left}
    \scedge{#1}{#2}{#4}{Right}{right}
}
\newcommand{\refts}[1]{Structure~T#1}
\newcommand{\reftsr}[1]{Rule~R#1}

\begin{document}
\title{Analogy-Making as a Core Primitive in the Software Engineering Toolbox}
\author{Matthew Sotoudeh}
\email{masotoudeh@ucdavis.edu}
\affiliation{%
  \institution{University of California, Davis}
}
\author{Aditya V. Thakur}
\email{avthakur@ucdavis.edu}
\affiliation{%
  \institution{University of California, Davis}
}

\renewcommand{\shortauthors}{Sotoudeh and Thakur}

\begin{abstract}
    An \emph{analogy} is an identification of structural similarities and
    correspondences between two objects. Computational models of analogy making
    have been studied extensively in the field of cognitive science to better
    understand high-level human cognition. For instance, Melanie Mitchell and
    Douglas Hofstadter sought to better understand high-level perception by
    developing the Copycat algorithm for completing analogies between letter
    sequences. In this paper, we argue that analogy making should be seen as a
    core primitive in software engineering. We motivate this argument by
    showing how complex software engineering problems such as program
    understanding and source-code transformation learning can be reduced to an
    instance of the analogy-making problem. We demonstrate this idea using
    \TC{}, a new analogy-making algorithm suitable for software engineering
    applications that adapts and extends ideas from Copycat. In particular,
    \TC{} reduces analogy-making to searching for a sequence of update rule
    applications. \TC{} uses a novel representation for mathematical structures
    capable of effectively representing the wide variety of information
    embedded in software. We conclude by listing major areas of future work for
    \TC{} and analogy-making in software engineering.
\end{abstract}

\maketitle

\section{Introduction}
An \emph{analogy} is defined as ``a comparison between two objects, or systems
of objects, that highlights respects in which they are thought to be
similar''~\cite{sep-reasoning-analogy}. Humans make complex, fluid analogies in
everyday communication. For example, a recent CNN headline \cite{cnn} states
that ``the `beating hearts' of these pulsating stars create music to
astronomers' ears,'' noting correspondences between pulsation of stars, the
beating of a heart, and the rhythm of music. In fact, analogy-making is such a
fundamental skill that a major portion of the original SAT exam was dedicated
to having test-takers complete analogies such as ``Paltry is to significance as
X is to Y.''

Analogies have been shown to be a useful instructional tool for improving
student learning \cite{gentner2003learning}. For instance, analogies are
frequently used to explain concepts in
physics~\cite{podolefsky2006use,clement1993using},
geography~\cite{nelson1975use}, mathematics~\cite{richland2007cognitive,richland2004analogy}, and
computer
science~\cite{douglas1983learning,san1993applying,DBLP:conf/ppig/Dunican02,forivsek2012metaphors,giacaman2012teaching,hidalgo2014playing,macfarlane1988study}.
For example, a professor of an introductory computer science course might
explain the concept of a program by forming an \emph{analogy between} the
execution of a program by a computer and the following of a recipe by a cook,
likening steps in a recipe to statements in a computer program, ingredients in a
recipe to resources in a program or user inputs --- analogies are not always
unambiguous.

The important role of analogy-making in high-level cognition has been argued by
many researchers in both cognitive science and computer
science~\cite{connectionistlogical,evans1964heuristic,falkenhainer1989structure}.
Professor Douglas Hofstadter and his Fluid Analogies Reasoning Group~(FARG)
\cite{DBLP:journals/aim/Hofstadter95,hofstadter1995fluid} argue that humans are
constantly making analogies, comparing features of their current situation with
previously-encountered scenarios to decide what to do next. To study the
analogy-making process in more detail, they have developed \emph{computational
models of analogy making.} One such algorithm,
\emph{Copycat}~\cite{hofstadter1994copycat}, can complete analogies over
strings~(\pref{sec:WhatIsAnalogyMaking}). Copycat can answer questions such as
``\stringobj{abc} is to \stringobj{abd} as \stringobj{efg} is to what?''

This paper explores the role of analogy-making in software engineering~(SE). We
argue that a number of SE problems can be framed as analogy-making. We introduce
a dedicated analogy-making algorithm, \TC{}, which can be used as a core
primitive to solve such SE problems. We believe that future research on
analogy-making algorithms like \TC{} will solidify analogy-making as a core
primitive in the software engineering toolbox.

The first application of analogy-making we consider is the problem of
\emph{program understanding}~(\pref{sec:ComparativeProgramUnderstanding}). One
fundamental way that humans understand new source code is by analogy to source
code that they already understand. As an example, consider a programmer who has
become familiar with the source code of the Bourne-Again Shell
(bash)~\cite{bash} and now wishes to add a new feature to the Friendly
Interactive Shell (fish)~\cite{fish}. A reasonable first step would be to read
the source code of fish and attempt to identify functions in its implementation
that play a similar role to more familiar functions in the implementation of
bash. In other words, the programmer will \emph{form an analogy} between the
two source code repositories and use the analogy to determine where to make the
desired modification.

The second application of analogy-making we consider is the problem of
\emph{generalizing a source code optimization}~(\pref{sec:WhatIfGEMM}).  As an
example, consider a scenario where an optimized matrix multiplication
implementation should be used when the matrix sizes satisfy certain conditions.
After seeing a small number of code corrections replacing the sub-optimal
multiplication routine with the optimized one, most trained programmers would
begin to internalize the pattern.  Consequently, when they come across new code
that can be optimized they would be able to form an analogy to those
corrections and modify the code to use the optimized version. Here, the
programmer needs to form an analogy \emph{between pairs of programs,} before
and after the transformation, for example stating that ``all of the sub-optimal
code calls matrix multiplication routine \stringobj{X} with inputs like
\stringobj{Y}, and the corresponding optimized code replaces \stringobj{X}{}
with \stringobj{Z}.''

Finally, the third application of analogy-making we consider is the problem of
\emph{API migration}~(\pref{sec:APIMigration}). As an example, consider the
scenario where a library that has updated its public interface to change its
error codes and remove a now-redundant parameter from each relevant function.
After updating a small number of functions to use the new API, a programmer
might quickly realize that all of the changes they need to make are
``fundamentally the same:'' lookup the new error code in the documentation,
switch the old error code for the new wherever it is checked for, and then
remove the redundant parameter. The programmer has, thus, formed an analogy
between the different edits, which they can use to quickly migrate similar code
elsewhere in the project.

In all of these examples, and many more discussed in~\pref{sec:WhatIfOther},
the programmer reasons about the source code and relations between different
parts of the source code to form an \emph{analogy} highlighting the fundamental
similarities between a set of examples.

Motivated by this, we developed \TC{}, an analogy-making algorithm suitable for
analogy-making on source code (\pref{sec:Implementation}). \TC{} takes as input
a description of the source code in a mathematical format using arbitrary
relations. This description is expressive enough to represent syntactic as well
as semantic information about the source code. \TC{} can also take as input and
reference in its analogies non-code sources like documentation. The design of
\TC{} is principled and ultimately reduced the analogy-making problem to that of
a search over possible rule applications for rewriting a workspace described as
a \emph{triplet structure}.

We believe that \TC{} can form a powerful primitive in the SE toolbox. \TC{} is
general enough to handle arbitrary relations, and, hence, can make analogies
synthesizing semantic, syntactic, and natural-language information. \TC{}'s
output is explainable, as it internally solves analogies by symbolic
manipulation and identifying corresponding facts. Finally, by reducing many
distinct problems to \TC{}, improvements to it will
directly pay dividends across a wide number of applications. 
The implementation of \TC{} is available at \url{https://github.com/95616ARG/sifter}.

\section{Analogy Making over Strings}
\label{sec:WhatIsAnalogyMaking}

This section explains the notion of analogy making.
Following Hofstadter, we focus first on analogies between \emph{letter
strings,} sequences of letters and symbols.
\pref{sec:LettersTransformations} then describes analogy completion.  We end by
describing Copycat (\pref{sec:Copycat}), a system from prior work for
completing analogies involving such letter strings, which influenced the design
of the \TC{} system introduced in this paper.

Informally, analogy making over strings entails determining in what respects
two given strings are similar. For instance, consider the strings
\stringobj{abc} and \stringobj{efg}. For the string \stringobj{abc}, we have
that the second letter is the successor in alphabetical order of the first, and
the third is the successor of the second. The exact same property holds for the
string \stringobj{efg}.

More formally, analogy making entails inferring properties that hold for both
strings. Consider the binary relation $\NextTo$, where $\NextTo(x,y)$ implies
that the letter $x$ is the left of the letter $y$ in a given string. For the
string \stringobj{abc}, we have $\NextTo(\stringobj{a},\stringobj{b})$ and
$\NextTo(\stringobj{b},\stringobj{c})$. Furthermore, consider the binary
relation $\LetterSuccessor$ where $\LetterSuccessor(x,y)$ holds if the letter
$y$ is the successor of letter $x$ in alphabetical order. In this example, we
have $\LetterSuccessor(\stringobj{a}, \stringobj{b})$,
$\LetterSuccessor(\stringobj{b},\stringobj{c})$,
$\LetterSuccessor(\stringobj{e},\stringobj{f})$, and so on. Using these two
relations, we can state that the elements $S = \{\stringobj{a}, \stringobj{b},
\stringobj{c} \}$ of the string \stringobj{abc} satisfy the following property
$\varphi$: $\forall  x, y \in S. \NextTo(x,y) \implies \LetterSuccessor(x,y)$.
We see that the elements of the string \stringobj{efg} satisfy this same
property. Hence we have formed a meaningful analogy between \stringobj{abc} and
\stringobj{efg}.

However, not all analogies can be succinctly expressed via a first-order logic
formula such as $\varphi$. Consider making an analogy between the strings
\stringobj{abc} and \stringobj{gfe}. In this case, the elements of
\stringobj{abc} satisfy $\varphi$, but the elements of \stringobj{gfe}
satisfy the property $\varphi'$: $\forall  x, y \in S.
\NextTo(x,y) \implies \LetterSuccessor(\mathbf{y,x})$. In particular, the order
of the arguments to $\LetterSuccessor$ in $\varphi$ and $\varphi'$ are
different. Intuitively, the property $\varphi'$ reads the string from right to
left, instead of left to right. Consequently, the analogy maker needs to be
able to express such ``slips:'' the strings \stringobj{abc} and \stringobj{gfe}
satisfy \emph{almost} the same property, \emph{except that} one reads the
string from left to right and the other from right to left.

Such properties can become even more difficult to state in a first-order logic
notation when \emph{grouping} is involved. Consider making an analogy between
the strings \stringobj{aaabbc} and \stringobj{ddddcccbba}. There are many
reasonable analogies between these strings. For example, we might associate the
\emph{group of letters} \stringobj{aaa} in the first string with \stringobj{a}
in the second string, noting that all of the letters in the first and the one
letter in the second satisfy the unary $\IsA$ predicate. Alternatively, we
might associate \stringobj{aaa} in the first string with \stringobj{dddd} in
the second string, as they are both groups of a single repeated letter
occurring at the start of their corresponding string.  In particular, the
strength of an analogy lies less in the number of features the two strings have
in common than in the overlap of relational structure between the two
strings~\cite{gentner1983structure}.  Regardless, it is not clear how one might
naturally express the fundamental properties that these two strings share in a
standard logic notation. Instead, as humans we might be tempted to communicate
the shared property via a drawing like the following:
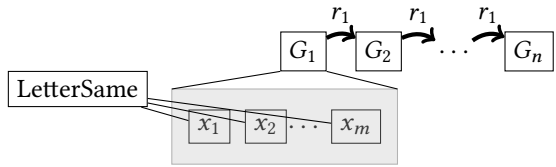
\begin{center}\begin{tikzpicture}
    \node[draw=black] (G1) at (0,0) {$G_1$};
    \node[draw=black] (G2) at (1,0) {$G_2$};
    \node (dots) at (2,0) {$\ldots$};
    \node[draw=black] (GN) at (3,0) {$G_n$};

    \node[draw=black] (X1) at (-1.25,-1) {$x_1$};
    \node[draw=black] (X2) at (-0.5,-1) {$x_2$};
    \node (dots2) at (0,-1) {$\ldots$};
    \node[draw=black] (XM) at (0.7,-1) {$x_m$};

    \node[draw=black] (Same) at (-3, -0.5) {$\LetterSame$};

    \draw[fill=lightgray,opacity=0.3] (-1.75, -1.5) rectangle (1.25, -0.5);
    \draw (G1.south west) -- (-1.75, -0.5);
    \draw (G1.south east) -- (1.25, -0.5);
    \draw (Same) -- (X1);
    \draw (Same) -- (X2);
    \draw (Same) -- (XM);

    \path[->,ultra thick] (G1) edge[bend left] node[above] {$r_1$} (G2);
    \path[->,ultra thick] (G2) edge[bend left] node[above] {$r_1$} (dots);
    \path[->,ultra thick] (dots) edge[bend left] node[above] {$r_1$} (GN);
\end{tikzpicture}\end{center}
In this drawing, $G_1$ through $G_n$ represent the groups of letters in each
string. Each group is made up of letters $x_1, \ldots, x_m$, which all
satisfy the $\LetterSame$ relation with each other. The groups themselves then
have some shared binary relation $r_1$ relating them. For example, in
\stringobj{aaabbc}, we might have \stringobj{aaa} take the place of $G_1$ and
$r_1(x, y) = \LetterSuccessor(x, y) \wedge \NextTo(x, y)$, while in
\stringobj{ddddcccbba} we might have \stringobj{a} take the place of $G_1$ with
$r_1(x, y) = \LetterSuccessor(x, y) \wedge \NextTo(\mathbf{y, x})$. Note again
that we have flipped the order in the latter $\NextTo$, intuitively to read the
second string from right-to-left.

Such drawings motivate an alternate way of thinking about analogy making. In
this interpretation, analogy-making involves defining an abstract string
description that can be instantiated to produce the given strings. As we have
seen, such an abstract string description needs to be general enough to handle
all of the complex grouping and slipping that can occur in such letter string
analogies.

\subsection{Analogy Completion on Transformations}
\label{sec:LettersTransformations}

One particularly interesting use-case for analogy-making is to make analogies
between \emph{pairs} of objects, such as a state before and after some
transformation. For example, given the pairs \stringpair{abc}{abd} and
\stringpair{efg}{efh}, we can compare the two pairs, forming an analogy which
might be represented by an abstraction like:
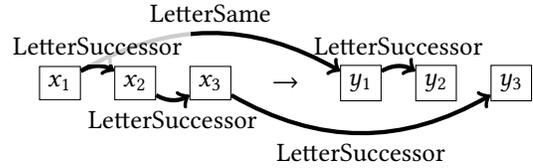
\begin{center}\begin{tikzpicture}
    \node[draw=black] (X1) at (0,0) {$x_1$};
    \node[draw=black] (X2) at (1,0) {$x_2$};
    \node[draw=black] (X3) at (2,0) {$x_3$};

    \node (Arrow) at (3,0) {$\to$};

    \node[draw=black] (Y1) at (4,0) {$y_1$};
    \node[draw=black] (Y2) at (5,0) {$y_2$};
    \node[draw=black] (Y3) at (6,0) {$y_3$};

    \path[->,ultra thick] (X1) edge[bend left] node[above] {$\LetterSame$} (Y1);

    \path[->,ultra thick] (X1) edge[bend left] node[above,fill=white,fill opacity=0.8,text opacity=1] {$\LetterSuccessor$} (X2);
    \path[->,ultra thick] (X2) edge[bend right] node[below] {$\LetterSuccessor$} (X3);

    \path[->,ultra thick] (Y1) edge[bend left] node[above] {$\LetterSuccessor$} (Y2);

    \path[->,ultra thick] (X3) edge[bend right] node[below] {$\LetterSuccessor$} (Y3);
\end{tikzpicture}\end{center}
In this scenario, the analogy-making process involves learning an
\emph{abstract representation} of the transformation performed in each pair of
strings. We have said that both pairs of strings correspond to each other
because they both share the properties shown in this diagram.

With such a representation of a transformation, one can also perform
\emph{Analogy Completion,} like in the original SAT exam. For example, given
two example pairs of strings \stringpair{abc}{abd} and \stringpair{efg}{efh}
and a \emph{prompt} \stringpair{ijk}{?}, we can ask for a \emph{completion of
the analogy,} a value which can replace the \stringobj{?} to make all three
pairs form a strong analogy. In this case, one completion would be the string
\stringobj{ijl}.

We can find such a completion by first constructing an analogy between the
examples \stringpair{abc}{abd} and \stringpair{efg}{efh} to form the abstract
string drawn above, then we can start to form an analogy with the examples and
the prompt \stringobj{ijk}, from which we might infer that \stringobj{i} is an
instance of $x_1$, \stringobj{j} an instance of $x_2$, and \stringobj{k} an
instance of $x_3$ in the abstract string. We can then \emph{infer} that there
should be some letters corresponding to $y_1, y_2, y_3$, and that they should
satisfy the properties in the drawing. From the drawing, then, we have
$\LetterSame(\stringobj{i}, y_1)$, hence we should have $y_1 = \stringobj{i}$;
then $\LetterSuccessor(y_1 = \stringobj{i}, y_2)$ to get $y_2 = \stringobj{j}$;
and finally $\LetterSuccessor(\stringobj{k}, y_3)$ to get $y_3 = \stringobj{l}$,
completing the analogy with the desired string \stringobj{ijl}.

\subsection{The Copycat Algorithm}
\label{sec:Copycat}

The Copycat algorithm \cite{hofstadter1994copycat} was developed to solve such
string analogy-completion problems. Copycat's architecture is similar to that of
a blackboard-based automated theorem prover. It consists of a \emph{workspace,}
or blackboard, which initially contains only the example pairs (such as
\stringpair{abc}{abd}) and a prompt (such as \stringpair{ijk}{?}). This
workspace is modified by a set of \emph{codelets}, which are small programs that
operate on the workspace. These codelets can make a variety of modifications to
the workspace. Some codelets may \emph{group} letters together, like the
\stringobj{aaa} in \stringobj{aaabbc}. Other codelets identify \emph{bonds}, or
relational facts about symbols, for example noting that
$\LetterSuccessor(\stringobj{a}, \stringobj{b})$ is true.  Still further
codelets can build \emph{bridges} between symbols, representing the determined
correspondences in the analogy. Once a consistent analogy among the examples and
prompt is made, a purpose-built solver is used to construct the corresponding
completion.

The resulting analogy is determined by the order and type of codelets used,
along with where each one ``focuses.'' The behavior of the codelets is
controlled by what is essentially a sophisticated set of heuristics wrapped
into a structure known as a \emph{Slipnet.}

\section{Applications of Analogy Makers in SE}
\label{sec:WhatIf}

In this section, we demonstrate the use of our analogy maker \TC{} by applying
it to three SE problems and list specific challenges addressed by $\TC$.  A
large number of further applications are then discussed
in~\pref{sec:WhatIfOther}.  We defer the description of the design of $\TC$ to
\pref{sec:Implementation}.

\subsection{Comparative Program Understanding}
\label{sec:ComparativeProgramUnderstanding}

\begin{figure*}[t]
    \begin{subfigure}{0.48\textwidth}
\begin{lstlisting}[frame=single,breaklines=true]
int cd_builtin (list) WORD_LIST *list; { ... } /*@\color{red}{//B1} @*/

struct builtin static_shell_builtins[] = { ...
  { "cd", cd_builtin, ... }, /*@\color{red}{//B2} @*/
... }
struct builtin *shell_builtins = static_shell_builtins; /*@\color{red}{//B3} @*/

struct builtin * builtin_address_internal
  (name, disabled_okay)
  char *name; int disabled_okay; { ... /*@\color{red}{//B4} @*/
    j = shell_builtins[mid].name[0] - name[0];
... }
\end{lstlisting}
        \caption{Bash source}
    \end{subfigure}
    \hfill
    \begin{subfigure}{0.48\textwidth}
\begin{lstlisting}[frame=single,breaklines=true]
int builtin_cd(parser_t &parser, io_streams_t &streams, wchar_t **argv) { ... } /*@\color{red}{//F1} @*/

static const builtin_data_t builtin_datas[] = {
...
  {L"cd", &builtin_cd, ...}, /*@\color{red}{//F2} @*/
... }

static const builtin_data_t *builtin_lookup(const wcstring &name) { /*@\color{red}{//F3} @*/
  const builtin_data_t *array_end = builtin_datas + BUILTIN_COUNT;
... }
\end{lstlisting}
        \caption{Fish source}
    \end{subfigure}
    \caption{Comparative program understanding between bash and fish source
    code. \TC{} can form an analogy between these implementations, helping
    explain the code for fish to a programmer used to the codebase of bash by
    noting where objects and functions (like
    \lstinline{builtin_address_internal} in bash and \lstinline{builtin_lookup}
    in fish) play similar roles in each.}
    \label{fig:SourceShell}
\end{figure*}

Suppose you are a programmer who is quite familiar with the implementation
(source code) of bash~\cite{bash}, and you would like to add a new feature to a
\emph{different} shell, such as fish~\cite{fish}. Because you have been adding
features to the bash shell for many years, you may know exactly which
function(s) to modify in the source code of bash to add the desired feature.
However, being new to the source code of fish, you face a significant challenge
understanding the implementation of fish before you can even begin writing your
new feature.

You might start by reading the source code of fish, looking for functions and
objects that play similar roles to ones you are more familiar with in bash.
This is fundamentally an analogy-making problem, where we are attempting to
form an analogy between the source code of bash (which we are familiar with) and
that of fish (which we are not).

This problem is similar to that of forming letter analogies
in~\pref{sec:WhatIsAnalogyMaking}, where, given the strings \stringobj{abc} and
\stringobj{efg}, we found that \stringobj{b} and \stringobj{f} corresponded to
each other. Here, \stringobj{abc} is instead the source code of bash, while
\stringobj{efg} is the source code of fish.  \stringobj{b} and \stringobj{f}
are likewise functions in bash and that which play similar roles.

\TC{} is designed to make such analogies on programs. We first load both source
repositories into \TC{}, then ask it to identify an analogy between the two.
This analogy will effectively be a \emph{mapping between the repositories},
identifying functions, classes, and statements in the source code of bash with
those in the source code of fish. Such an analogy can form an invaluable guide,
allowing you to look up, for example, the function in fish which \TC{} thinks
plays the most similar role to the one you would have modified in bash.

\subsubsection*{Challenge 1: Compositional Structure of Programs}

Program source-code relies heavily on compositional structure. For example, the
meaning and interpretation of any function depends not only on its immediate
body, but also on the body of all functions that it calls, and the functions
they call, and so forth. On the other hand, the role a function plays in a
large piece of software usually depends on the functions that call it.

\TC{} can make use of this compositional structure, by building on top of
analogies it has already made. For example, in~\pref{fig:SourceShell} we have
provided snippets of the source code of the bash and fish shells. \TC{} begins
by associating the functions \lstinline{cd_builtin} (\labeltext{B1}) and
\lstinline{builtin_cd} (\labeltext{F1}) because of their similar names and signatures.
Then, once it has decided that these two functions correspond, \TC{} can start
to make inferences that the places where they are used are likely to correspond
as well. For example, it might note that both \lstinline{shell_builtins} (\labeltext{B3})
and \lstinline{builtin_datas} (\labeltext{F2}) contain a struct with a field of
\lstinline{cd_builtin} or \lstinline{builtin_cd} respectively, and mark those
two objects as corresponding in the analogy. It can similarly infer that
functions using those objects, such as \lstinline{builtin_address_internal}
(\labeltext{B4}) in bash and \lstinline{builtin_lookup} (\labeltext{F3}) in fish, correspond.

In this way, \TC{} can build up analogies made about parts of the program
to begin to make stronger and stronger inferences about how the rest of the
source code corresponds. Although we have not demonstrated it in this example,
\TC{} can also make analogies in a top-down fashion, e.g., by starting at the
\lstinline{main} function in both programs, or by
alternating between such top-down and bottom-up strategies. Notably, such
compositional structure was not needed to find analogies between the letter
groups in~\pref{sec:WhatIsAnalogyMaking}, demonstrating how analogy-making on
programs can be richer and more challenging than on letter strings.

\subsubsection*{Challenge 2: Multiple Syntactic Representations of Programs}

Another challenge with analogy-making on programs is that semantically equivalent programs can
have multiple syntactic representations. For example, functions can be inlined
or \lstinline{if/else} conditionals can be inverted. \TC{} can handle such
scenarios by applying \emph{transformation rules,} such as function inlining,
to transform either source repository it is given. \TC{} searches through
different representations of each source repository until it finds ones that
are amenable to forming strong analogies.

Because letter strings do not have an assumed semantics, there is no equivalent
notion of semantics-preserving transformation rules for the examples
in~\pref{sec:WhatIsAnalogyMaking}. However, operationally, the process of
grouping letters, e.g., in \stringobj{aaabbc}, can be seen as such a
transformation, where the internal representation of the individual letters
\stringobj{a}, \stringobj{a}, and \stringobj{a} are transformed into a single
\emph{group} of letters \stringobj{aaa}.

\subsection{Generalizing Program Transformations from Examples}
\label{sec:WhatIfGEMM}

Suppose we have a linear algebra library with multiple General Matrix Multiply (GEMM) routines for
computing matrix multiplications. Some routines, such as
\lstinline{gemm_large}, are optimized for the case where the input matrices are
relatively large, say with over 1,000 rows each, while others like
\lstinline{gemm_skinny} are optimized for ``skinny'' inputs, e.g., where the
inner dimension is half the size of either of the outer dimensions.

For a particular team working on a particular codebase, it may be the case that
most matrices are usually quite large and so \lstinline{gemm_large} might become
the de-facto routine that developers use in new code without thinking too deeply
about matrix sizes, or simply used due to copy and paste from existing code.
While this might be a reasonable default for this team, it is likely the case
that \emph{some} matrices in a program are better suited for
\lstinline{gemm_skinny} --- in that case, the instinctive default would be
sub-optimal, and another programmer might notice during code review that
\lstinline{gemm_skinny} would be a better choice.

The question we would like to consider is: given two example code pairs where
\lstinline{gemm_large} has been transformed into \lstinline{gemm_skinny}, can
we automatically optimize new code in the same manner? This is fundamentally an
analogy problem, where we would like to compare the pairs of pre- and
post-replacement code
to learn the \emph{core transformation} that explains all of them. We can then
use this analogy to infer, for some new sub-optimal code, the corresponding
optimized code. This is similar to completing the analogy \stringpair{abc}{abd},
\stringpair{efg}{efh}, \stringpair{ijk}{?}
in~\pref{sec:LettersTransformations}.

This scenario is shown in~\pref{fig:GEMM}. The first two rows in that figure
show pairs of examples of the desired source code transformation provided to
\TC{}, which play the same role as \stringpair{abc}{abd} and
\stringpair{efg}{efh} in~\pref{sec:LettersTransformations}.  The code in the
left-hand column is sub-optimal because it calls \lstinline{gemm_large} on
matrices with dimensions that would be better suited for
\lstinline{gemm_skinny}. The code in the right-hand column has been optimized
by replacing the call to \lstinline{gemm_large} with a call to
\lstinline{gemm_skinny}. In the third row of~\pref{fig:GEMM}, we have provided
\TC{} with a new piece of code on the left and ask it to \emph{complete the
analogy,} i.e., produce the corresponding piece of code on the right that
makes all three rows the most similar. This plays the same role as the
\stringpair{efg}{?} input in the letter analogy example. The code produced by
\TC{} is shown on the bottom right of~\pref{fig:GEMM} in green, where we see it
has correctly replaced the call to \lstinline{gemm_large} with a call to
\lstinline{gemm_skinny}.

\begin{figure*}
    \begin{subfigure}{0.48\textwidth}
\begin{lstlisting}[frame=single,breaklines=true]
assert(k > 0);
int outer = k * 100;
int inner = k * 10;
read_mat(outer, inner, &A);
read_mat(inner, outer, &B);
gemm_large(A, B, &C, outer, inner, outer);
\end{lstlisting}
    \end{subfigure}
    \hfill
    \begin{subfigure}{0.48\textwidth}
\begin{lstlisting}[frame=single,breaklines=true]
assert(k > 0);
int outer = k * 100;
int inner = k * 10;
read_mat(outer, inner, &A);
read_mat(inner, outer, &B);
gemm_skinny(A, B, &C, outer, inner, outer);
\end{lstlisting}
    \end{subfigure}
    \begin{subfigure}{0.48\textwidth}
\begin{lstlisting}[frame=single,breaklines=true]
assert(k > 1);
int outer = k, A_cols = k / 2;
read_mat(outer, A_cols, &A);
read_mat(A_cols, outer, &B);
while (!done(A, B)) {
    read_row(&A);
    read_col(&B);
    outer++; }
gemm_large(A, B, &C, outer, A_cols, outer);
\end{lstlisting}
    \end{subfigure}
    \hfill
    \begin{subfigure}{0.48\textwidth}
\begin{lstlisting}[frame=single,breaklines=true]
assert(k > 1);
int outer = k, A_cols = k / 2;
read_mat(outer, A_cols, &A);
read_mat(A_cols, outer, &B);
while (!done(A, B)) {
    read_row(&A);
    read_col(&B);
    outer++; }
gemm_skinny(A, B, &C, outer, A_cols, outer);
\end{lstlisting}
    \end{subfigure}
    \begin{subfigure}{0.48\textwidth}
\begin{lstlisting}[frame=single,breaklines=true]
assert(k > 5);
int AB_rowcol = k * k;
int inner = k;
read_mat(AB_rowcol, inner, &A);
read_mat(inner, AB_rowcol, &B);
gemm_large(A,B,&C, AB_rowcol, inner, AB_rowcol);
\end{lstlisting}
    \end{subfigure}
    \hfill
    \begin{subfigure}{0.48\textwidth}
\begin{lstlisting}[frame=single,breaklines=true,backgroundcolor=\color{green}]
assert(k > 5);
int AB_rowcol = k * k;
int inner = k;
read_mat(AB_rowcol, inner, &A);
read_mat(inner, AB_rowcol, &B);
gemm_skinny(A,B,&C, AB_rowcol, inner, AB_rowcol);
\end{lstlisting}
    \end{subfigure}
    \caption{Optimizing program source code with \TC{}. The left column shows
    before the optimization and the right column after the optimization. The
    first two lines are the examples given to \TC{}, while the green code is
    generated by \TC{} to complete the analogy for the last line. \TC{}
    includes in its analogy semantic information like the fact that all of the
    outer dimensions in the examples are at least twice that of the inner
    dimensions, helping it to avoid false positives.}
    \label{fig:GEMM}
\end{figure*}

\subsubsection*{Challenge 3: Avoiding False Positives}

\TC{} is forming an analogy between the rows in~\pref{fig:GEMM},
including between the before code on the left-hand side. In~\pref{fig:GEMM},
for example, \TC{} has noted as part of its analogy that all of the left-hand
code snippets call the function \lstinline{gemm_large} with the last argument
at least twice that of the second-to-last argument. This behavior can be
thought of as learning to \emph{recognize code that can be optimized}, and can
be used to avoid false positives.  For example, suppose instead of the
sub-optimal prompt code given in the bottom-left of~\pref{fig:GEMM}, we gave $\TC$
the code: 
\begin{lstlisting}[frame=single,breaklines=true]
assert(k > 0);
int outer = k * 10;
int inner = k * 10;
read_mat(outer, inner, &A);
read_mat(inner, outer, &B);
gemm_large(A, B, &C, outer, inner, outer);
\end{lstlisting}
In that scenario, \TC{} would attempt to form an analogy between this code and
the example before-transformation code on the left-hand side of the first two
rows of~\pref{fig:GEMM}. While it may succeed in proposing an analogy, \TC{}
will note as part of its output that the analogy is not particularly strong.
This is because this new code does not share the property that the last
argument to \lstinline{gemm_large} is at least twice that of the second-to-last
one.

If examples of already-optimized code are available, given a new instance \TC{}
can also try to form an analogy using these \emph{negative} examples. A
threshold can be set based on a comparison with the negative vs.\ positive
examples to determine whether to apply the transformation.

\subsubsection*{Challenge 4: Using Semantic Information}

Recognizing sub-optimal code relies on \emph{semantic information} about the
possible values a variable can take on. In particular, we only want to apply the
transformation when the inner matrix dimension is at most half the size of the
outer dimensions. This would cause difficulty for syntax-based tools like
GetAFix~\cite{bader2019getafix}. However, \TC{} takes as input an
\emph{arbitrary structure} consisting of symbols and relations between the
symbols. This means that, in addition to providing the source code, we can
annotate the structure representing the source code with the results of a
program analyzer, which allows us to include information about semantic
properties of the code. In this example, we can annotate the structure with
relations that some variable is always at least twice that of another. The
analogy-making algorithm that forms the core of \TC{} is entirely indifferent to
the underlying relations being used, and will use such semantic relations to
form analogies just like it would more syntactic relationships describing the
source code.

\subsection{API Migration}
\label{sec:APIMigration}

\begin{figure*}
    \begin{subfigure}{0.48\textwidth}
\begin{lstlisting}[frame=single,breaklines=true]
# CameraLib v1.0
## `record_video(buffer, buffer_size, resolution)`
Records video from the main camera into `buffer` until `buffer_size` bytes are reached. On error returns -1.
## `record_audio(buffer, buffer_size, resolution)`
Uses the main camera's microphone to record audio into `buffer` until `buffer_size` bytes have been recorded. On error returns -5.
## `record_frame(buffer, buffer_size, resolution)`
Uses the main camera to record a single image to `buffer`. On failure returns -3.
\end{lstlisting}
    \end{subfigure}
    \hfill
    \begin{subfigure}{0.48\textwidth}
\begin{lstlisting}[frame=single,breaklines=true]
# CameraLib v2.0
## `record_video(buffer, buffer_size)`
Records video from the main camera into `buffer` until `buffer_size` bytes are reached. On error returns -4.
## `record_audio(buffer, buffer_size)`
Uses the main camera's microphone to record audio into `buffer` until `buffer_size` bytes have been recorded. On error returns -2.
## `record_frame(buffer, buffer_size)`
Uses the main camera to record a single image to `buffer`. Automatically sets the resolution to fit in `buffer_size`. On failure returns -6.
\end{lstlisting}
    \end{subfigure}
    \caption{API documentation before after migration for the camera API. This
    documentation is provided to \TC{} to complete the analogy shown
    in~\pref{fig:API}.}
    \label{fig:APIDocs}
\end{figure*}

For the final task, consider the two versions of a camera library documented
in~\pref{fig:APIDocs}, which has been updated to automatically determine the
resolution to use as well as changed the error codes. Suppose you have a program
that uses version 1 of this camera library.  Snippets of this program are
shown on the left column of~\pref{fig:API}.  Your
colleague has partially migrated your code to use version 2
of the camera library, as shown in the first two boxes on the right column
of~\pref{fig:API}. You would now like to migrate the bottom-left code
in~\pref{fig:API} to use version 2.

This is an analogy completion problem, where we want to form an analogy
between the examples that effectively describes how to migrate code to use the
new API. We then want to find a \emph{completion}, or source code to fill in to
the bottom-right of~\pref{fig:API} that makes the last row analogous to the
first. Because this involves transforming source code, the input/output to
\TC{} will be similar to that of~\pref{sec:WhatIfGEMM}. However, as we will
see, we will face different challenges here.

\begin{figure*}
    \begin{subfigure}{0.48\textwidth}
\begin{lstlisting}
...
void try_record_video() {
  int result = record_video(buffer, BUFFER_SIZE, RES_AUTO);
  if (result == -1) {
    printf("Could not record video.\n"); }
... }
\end{lstlisting}
    \end{subfigure}
    \hfill
    \begin{subfigure}{0.48\textwidth}
\begin{lstlisting}[frame=single,breaklines=true]
...
void try_record_video() {
  int result = record_video(buffer, BUFFER_SIZE);
  if (result == -4) {
    printf("Could not record video.\n"); }
... }
\end{lstlisting}
    \end{subfigure}
    \begin{subfigure}{0.48\textwidth}
\begin{lstlisting}[frame=single,breaklines=true]
...
void try_record_audio() {
  int result = record_audio(buffer, BUFFER_SIZE, RES_AUTO);
  if (result == -5) {
    printf("Could not record audio.\n"); }
... }
\end{lstlisting}
    \end{subfigure}
    \hfill
    \begin{subfigure}{0.48\textwidth}
\begin{lstlisting}[frame=single,breaklines=true]
...
void try_record_audio() {
  int result = record_audio(buffer, BUFFER_SIZE);
  if (result == -2) {
    printf("Could not record audio.\n"); }
... }
\end{lstlisting}
    \end{subfigure}
    \begin{subfigure}{0.48\textwidth}
\begin{lstlisting}[frame=single,breaklines=true]
...
void try_record_still() {
  int result = record_frame(buffer, BUFFER_SIZE, RES_AUTO);
  if (result == -3) {
    printf("Could not record still.\n"); }
... }
\end{lstlisting}
    \end{subfigure}
    \hfill
    \begin{subfigure}{0.48\textwidth}
\begin{lstlisting}[frame=single,breaklines=true,backgroundcolor=\color{green}]
...
void try_record_still() {
  int result = record_frame(buffer, BUFFER_SIZE);
  if (result == -6) {
    printf("Could not record still.\n"); }
... }
\end{lstlisting}
    \end{subfigure}
    \caption{API migration with \TC{}. The left (right) column shows before
    (after) the migration. The first two rows are the input to \TC{},
    while the green code is generated by \TC{} to complete the analogy for the
    last row.  \TC{} was also given as input the API documentation pair from
    \pref{fig:APIDocs}, which is where it looks to find the new error code used
    in the generated code.
    }
    \label{fig:API}
\end{figure*}

Just like in~\pref{sec:WhatIfGEMM}, we can provide the first two rows
of~\pref{fig:API} to \TC{} as examples of the desired migration, then ask it to
\emph{complete the analogy} by producing the migrated code corresponding to the
prompt (unmigrated) code in the bottom left of~\pref{fig:API}. The code
generated by \TC{} is shown in green on the bottom left of~\pref{fig:API}.

\subsubsection*{Challenge 5: Using Documentation in Analogies}
There is one particularly pressing challenge we wish to highlight here: given
\emph{only} these code pairs, there is no reasonable way to complete this
analogy, because it depends on knowing the new error code for
\lstinline{record_frame}, which did not appear in any of the examples. To
address this, we can \emph{give \TC{} the before/after documentation} in
addition to just the source code.  Again, because \TC{} takes any structure as
input, we can encode documentation just as easily as we can encode source code.
Then, in making analogies, \TC{} can \emph{refer to the documentation} and
include looking up in the documentation as part of the analogy.

\subsubsection*{Challenge 6: Using DNN Models in Analogy-Making}
Once we begin involving arbitrary text (e.g., in documentation), we
need to start being able to handle fuzziness inherent in human languages. For
example, in the documentation for the first two functions, the corresponding
error message could be found with a relatively simple search, because it was
prefaced with ``On error \ldots returns.'' However, for the
\lstinline{record_frame} function, the corresponding sentence uses ``failure''
instead of ``error,'' which could cause \TC{} to lose confidence in its
analogy. To help increase \TC{}'s confidence in its analogy and guide it
towards the right answer, we can use existing natural-language tools such as
DNN-based sentiment analysis models. These models take a paragraph, sentence,
or word and estimate how positive or negative it is. Just like with the results
of a program analyzer, we can annotate this information on top of the
structure, e.g., by marking all symbols representing words which are determined
to be highly positive or highly negative by the sentiment analyzer with a unary
relation like $\IsNegativeSentiment$. Even though they are not exactly the same
word, the fact that they both satisfy the $\IsNegativeSentiment$ relation will
increase \TC{}'s confidence in the analogy.

\subsection{Future Software Engineering Applications}
\label{sec:WhatIfOther}
Future work can apply analogy completion to more varied input/output domains
and languages.  For example, analogy completion can \emph{generate
documentation} based on existing code, similar to the problem of
automated comment generation, which has been addressed by code-clone
detection~\cite{wong2015clocom}. We can also treat translation between
programming languages as analogy completion using examples of semantically
equivalent programs in either language. This will likely rely on a number of
smaller analogies, matching common patterns in one programming language and
mapping them to idiomatic code for that pattern in another, similar to existing
work in this field~\cite{miltner2017synthesizing,maina2018synthesizing}.

A smart editor may make analogies between the user's current editor state and a
corpus of code samples to suggest structural code completions, similar to those
accomplished using large code corpuses like Aroma~\cite{luan2019aroma} or more
local history such as Blue~Pencil~\cite{miltner:blue_pencil}.  Because our
approach does not rely on AST parsing (see~\pref{sec:RepresentingCode}), we can
use analogies to automatically improve tooling (error messages, linting,
suggestions, bug finding, etc.) for nascent and domain-specific languages
(DSLs) that may not yet have a formal
grammar~\cite{nilsson2008practical,microgrammars}.

One might also apply analogy-makers to recover higher-level structure from
low-level binaries or compiler intermediate representations. Such information
would be useful for decompilation of binary programs~\cite{andriesse2016depth},
in-place binary analysis~\cite{song2008bitblaze}, binary
rewriting~\cite{wenzl2019hack}, de-obfuscation of
code~\cite{raychev2015predicting}, and identification of code replacable with
highly-optimized libraries~\cite{di2014towards} or
coprocessors~\cite{sotoudeh2019isa}.

Analogies between existing code with correctness proofs and new code may allow
for \emph{proof transfer} to more quickly prove correctness of the latter.  A
related technique has been proposed for the Coq theorem
prover~\cite{pfenning1991unification}.

More varied information sources can be used for analogy making. For example,
the use of information from a profiler may be helpful in a code optimization
setting, and compiler error-messages may be useful when using analogies to
provide edit suggestions for syntactically-invalid code. This information could
be used to rank different proposed analogies or to help find them in the first
place.

In a classroom setting, analogy-makers can cluster student assignments, an
important problem which is currently addressed via a variety of different
techniques~\cite{fse:studentfeedback,gupta2019deep}.

\TC{} can also detect where a strong analogy \emph{almost holds,} under a small
modification its inputs, e.g., in \stringobj{abc} and \stringobj{xyf}. If
applied to common coding patterns, this might suggest the existence of a bug in
the program's implementation of this pattern.  For example, one binary-search
implementation might be \emph{almost} analogous to a reference one, except that
it computes the midpoint as \lstinline{(l+h)/2} instead of
\lstinline{l+(h-l)/2}, introducing a subtle integer overflow bug that a future
version of \TC{} might flag as anomalous in the analogy.

\section{Design of \TC{}}
\label{sec:Implementation}

In this section, we describe the design of our analogy-making algorithm $\TC$
and illustrate how it addresses the challenges discussed
in~\pref{sec:WhatIf}. Its design was influenced by that of
Copycat~(\pref{sec:Copycat}). However, as we will discuss
in~\pref{sec:RelatedWork}, Copycat's implementation was specially designed for
the letter-analogy domain, whereas we would like to support \emph{arbitrary}
relations and input structures.

At a high level, the behavior of \TC{} is formulated as a number of
\emph{update rules} operating on a \emph{workspace.} The workspace initially
contains a representation of the source code and other inputs which it is
supposed to make analogies about. Update rules gradually modify the workspace,
both identifying facts, such as when some letter is a successor of another, and
making new analogies. Analogies made during this process are explicitly
represented within the workspace, as discussed
in~\pref{sec:RepresentingAnalogies}, and new analogies can build iteratively on
existing analogies in the workspace. When a sufficient analogy is found by the
system, it can be read off directly from the workspace and returned to the
user.

In~\pref{sec:ImplementationTripletStructures} we will describe \emph{triplet
structures,} a novel data structure used to represent \TC{}'s
workspace. Triplet structures can represent arbitrary relational
facts in a standardized way, making them a particularly flexible tool for
representing \TC{}'s workspace. In~\pref{sec:InitWorkspace} we will describe
how we initialize the triplet structure representing the \TC{}
workspace for an example analogy problem. In~\pref{sec:RewriteRules}, we
introduce a domain-specific language for expressing \emph{update rules} that
modify triplet structures, and can be used to
infer new facts about the objects in question.
\pref{sec:RepresentingAnalogies} describes how analogies are represented in the
workspace, while~\pref{sec:ImplementationAnalogyRules} describes update rules
that can be used to automatically find such analogies.

\subsection{Triplet Structures}
\label{sec:ImplementationTripletStructures}
A \emph{triplet structure}
is a novel data structure used to represent the state of
\TC{}'s workspace. A triplet structure:
\begin{enumerate}
    \item Represents objects and facts in a standardized form, so that code for
        operating on the workspace does not have to worry about details like
        arity of relations.
    \item Is able to naturally represent \emph{partial facts,} e.g., we can
        represent the state ``I know letter O is the predecessor of
        \emph{something,} but I'm not sure exactly what yet.''
    \item Supports efficient lookups and queries, so that operations on the
        structure can be performed quickly.
\end{enumerate}

\begin{definition}
    A \emph{triplet structure} is a pair of sets $(S, F)$ where $F \subseteq
    S\times S\times S$. We call each member of $S$ a \emph{node,} each member
    of $S\times S \times S$ a \emph{triplet fact,} and $F$ the set of triplet
    facts \emph{in the structure.}
\end{definition}

We can encode any finite mathematical structure as a triplet structure with
polynomial increase in size. First, for every $n$-ary relation $R$, we add $n$
nodes to the triplet structure representing slots in the relation. Generally,
for an $n$-ary relation $R$ we can always add nodes $\ROne$ through $\Rn$,
although we will usually use more descriptive names in our examples.  Second,
each fact in the original structure gets a \emph{fact node} in the triplet
structure, which is a node in $S$ that \emph{represents
the original fact itself} in the triplet structure. For a fact in the original
structure of the form $R(x_1, x_2, \ldots, x_n)$ corresponding to a fact node
$f$, we then add triplet facts of the form $(f, x_i, \Ri)$ for every $i \in \{
    1, 2, \ldots, n \}$.

Fact nodes in triplet structures can be thought of as C-style
\lstinline{struct}s, where each fact $(f, v, k)$ asserts that the field $k$ in
struct $f$ takes the value $v$. Alternatively, each fact node~$f$ can be
thought of as expressing an interpretation of part of the structure, with a
fact $(f, v, k)$ asserting that, in the interpretation $f$, $v$ is of type $k$.

\begin{example}
    \label{ex:TripletStructure}
    Consider a mathematical structure with objects $O = \{ x, y, z \}$, a
    single binary relation $R$, and two relational facts $R(x, y)$, $R(y, z)$.

    To encode this mathematical structure as a
    triplet structure, we break the binary relation $R$ into two nodes $\ROne$
    and $\RTwo$ representing each of its slots. We then create the fact
    node~$f_1$ for $R(x, y)$ and the fact node $f_2$ for $R(y, z)$. We also add
    nodes for each of the original objects in $O$ to get the set of nodes:
    \[
        S = \{ x, y, z, \ROne, \RTwo, f_1, f_2 \}.
    \]
    Finally, we add triplet facts relating each slot of each fact to arrive at
    the set of triplet facts in the structure:
    \[
        \begin{aligned}
            F = \{
                &(f_1, x, \ROne), \quad (f_1, y, \RTwo), \\
                &(f_2, y, \ROne), \quad (f_2, z, \RTwo) \}.
        \end{aligned}
    \]

    Triplet structures also have an intuitive graph representation.  Nodes in
    the structure correspond to nodes in the graph.  For each triplet fact $(f,
    v, k)$, we add an edge $v\to k$ with label~$f$.  The graph for the triplet
    structure considered in this example is the~\refts{1} shown
    below.
    \begin{center} \begin{tikzpicture}
        \node[circle,draw=black] (A) at (-2,0) {$x$};
        \node[circle,draw=black] (B) at (0,0)  {$y$};
        \node[circle,draw=black] (C) at (2,0)  {$z$};
        \node[circle,fill=red] (F1) at (4,1) {$f_1$};
        \node[circle,fill=orange] (F2) at (4,0) {$f_2$};
        \node[draw=black] (L) at (-2,1) {$\ROne$};
        \node[draw=black] (R) at (2,1) {$\RTwo$};

        \path[->,ultra thick,draw=red] (A) edge node[left] {$f_1$} (L);
        \path[->,ultra thick,draw=red] (B) edge node[below] {$f_1$} (R);
        \path[->,ultra thick,draw=orange] (B) edge node[below] {$f_2$} (L);
        \path[->,ultra thick,draw=orange] (C) edge node[right] {$f_2$} (R);

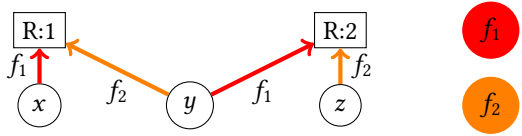
        \node (label) at (0, -1) {\textbf{Triplet Structure T1}};
    \end{tikzpicture} \end{center}
    Note that, while we have drawn T1 using descriptive names, shapes, and
    colors, no
    intrinsic meaning is assigned to any symbol. In the rest of this paper, we
    will usually only show this visual representation of a triplet structure
    instead of explicitly listing the nodes and facts. Hence, the reader is
    encouraged to ensure the connection between the two is well-understood
    before proceeding.
\end{example}

In addition to directly encoding relational facts,
some structures can be more naturally expressed directly as a triplet
structure. This is highlighted in the next example.
\begin{example}
    Consider encoding the scenario ``Homer and Marge are the parents of Bart
    and Lisa.'' We may encode this
    as four facts of the form $Parent(Homer, Bart)$,
    \\
    $Parent(Homer, Lisa)$,
    $Parent(Marge, Bart)$, $Parent(Marge, Lisa)$. With triplet structures, we can
    express this by saying ``Homer, Marge, Bart, and Lisa form a
    family, where Homer and Marge are the parents, and Bart and Lisa are the
    children.'' This scenario is represented by~\refts{2} below.
    \begin{center} \begin{tikzpicture}
        \node[draw=black] (H) at (-3,0) {Homer};
        \node[draw=black] (M) at (-1,0)  {Marge};
        \node[draw=black] (B) at (1,0)  {Bart};
        \node[draw=black] (L) at (3,0)  {Lisa};
        \node[circle,fill=red] (F1) at (0,-1) {$f_1$};
        \node[draw=black] (P) at (-2,1.5) {Family:Parents};
        \node[draw=black] (C) at (2,1.5) {Family:Children};

        \path[->,ultra thick,draw=red] (H) edge node[left] {$f_1$} (P);
        \path[->,ultra thick,draw=red] (M) edge node[right] {$f_1$} (P);
        \path[->,ultra thick,draw=red] (B) edge node[left] {$f_1$} (C);
        \path[->,ultra thick,draw=red] (L) edge node[right] {$f_1$} (C);
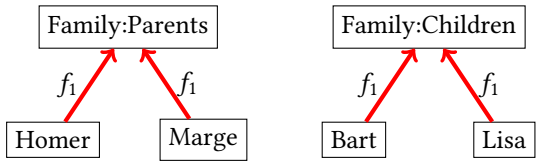
        \node (label) at (0, -2) {\textbf{Triplet Structure T2}};
    \end{tikzpicture} \end{center}
\end{example}

Another feature of triplet structures is that they can represent \emph{partial
facts}, as demonstrated by the next example.
\begin{example}
    Suppose in the previous example that we know Abe is the parent
    of \emph{someone,} but we are not sure who yet. We represent this
    uncertainty in~\refts{3} below by adding a new fact node $f_2$,
    which only states that Abe is a parent, without noting a corresponding
    child.
    \begin{center} \begin{tikzpicture}
        \node[draw=black] (A) at (-3.5,0) {Abe};
        \node[draw=black] (H) at (-2,0) {Homer};
        \node[draw=black] (M) at (-0.5,0)  {Marge};
        \node[draw=black] (B) at (1,0)  {Bart};
        \node[draw=black] (L) at (3,0)  {Lisa};
        \node[circle,fill=red] (F1) at (-1,-1) {$f_1$};
        \node[circle,fill=green] (F2) at (1,-1) {$f_2$};
        \node[draw=black] (P) at (-2,1.5) {Family:Parents};
        \node[draw=black] (C) at (2,1.5) {Family:Children};

        \path[->,ultra thick,draw=red] (H) edge node[left] {$f_1$} (P);
        \path[->,ultra thick,draw=red] (M) edge node[right] {$f_1$} (P);
        \path[->,ultra thick,draw=red] (B) edge node[left] {$f_1$} (C);
        \path[->,ultra thick,draw=red] (L) edge node[right] {$f_1$} (C);
        \path[->,ultra thick,draw=green] (A) edge node[left] {$f_2$} (P);
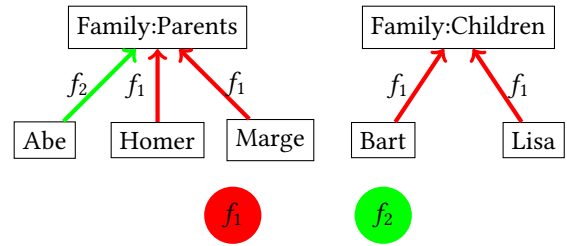
        \node (label) at (0, -2) {\textbf{Triplet Structure T3}};
    \end{tikzpicture} \end{center}
    If we later learn that Homer is Abe's child, we can extend $f_2$ to include
    this information as shown in~\refts{4} below.
    \begin{center} \begin{tikzpicture}
        \node[draw=black] (A) at (-3.5,0) {Abe};
        \node[draw=black] (H) at (-2,0) {Homer};
        \node[draw=black] (M) at (-0.5,0)  {Marge};
        \node[draw=black] (B) at (1,0)  {Bart};
        \node[draw=black] (L) at (3,0)  {Lisa};
        \node[circle,fill=red] (F1) at (-1,-1) {$f_1$};
        \node[circle,fill=green] (F2) at (1,-1) {$f_2$};
        \node[draw=black] (P) at (-2,1.5) {Family:Parents};
        \node[draw=black] (C) at (2,1.5) {Family:Children};

        \path[->,ultra thick,draw=red] (H) edge node[left] {$f_1$} (P);
        \path[->,ultra thick,draw=red] (M) edge node[right] {$f_1$} (P);
        \path[->,ultra thick,draw=red] (B) edge node[left] {$f_1$} (C);
        \path[->,ultra thick,draw=red] (L) edge node[right] {$f_1$} (C);
        \path[->,ultra thick,draw=green] (A) edge node[left] {$f_2$} (P);
        \path[->,ultra thick,draw=green] (H) edge node[above,near end,xshift=-2mm,yshift=-2mm] {$f_2$} (C);
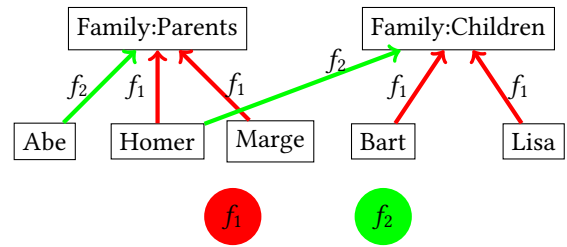
        \node (label) at (0, -2) {\textbf{Triplet Structure T4}};
    \end{tikzpicture} \end{center}
\end{example}

\newcommand{\tikzruleall}[2]{
\begin{center} \begin{tikzpicture}[scale=0.87]
    \node[circle,draw=black] (A) at (-2,0) {$x_1$};
    \node[circle,draw=black] (B) at (-1,0) {$x_2$};
    \node[circle,draw=black] (E) at (1,0)  {$y_1$};
    \node[circle,draw=black] (F) at (2,0)  {$y_2$};

    \node[draw=black] (PA) at (-4,0)  {\PlatonicLetter{a}};
    \node[draw=black] (PB) at (-4,-1) {\PlatonicLetter{b}};
    \node[draw=black] (PF) at (4,0)   {\PlatonicLetter{f}};
    \node[draw=black] (PE) at (4,-1)  {\PlatonicLetter{e}};

    \ifthenelse{\equal{#1}{before}}{
        \node[draw=black] (Pred) at (-2,1) {Predecessor};
        \node[draw=black] (Succ) at (2, 1)  {Successor};
    }{
        \node[draw=black] (Pred) at (-2,2) {Predecessor};
        \node[draw=black] (Succ) at (2, 2)  {Successor};
    }

    \node[draw=black] (Left) at (-2,-2) {$\NextToLeft$};
    \node[draw=black] (Right) at (2,-2) {$\NextToRight$};

    \node[circle,fill=red,draw=black] (MN1) at (-2,-3) {$n_1$};
    \node[circle,fill=orange,draw=black] (MN2) at (0.75,-3) {$n_2$};
    \ifthenelse{\equal{#1}{before}}{
    }{
        \node[circle,fill=cyan,draw=black] (MS1) at (-0.75,-3) {$s_1$};
    }
    \ifthenelse{\equal{#1}{after2}}{
        \node[circle,fill=green,draw=black] (MS2) at (2,-3) {$s_2$};
    }{
    }

    \node[circle,fill=lightgray,draw=black] (MP1) at (-4,-3) {$p_1$};
    \node[circle,fill=purple,draw=black] (MP2) at    (-3,-3) {$p_2$};
    \node[circle,fill=pink,draw=black] (MP3) at       (3,-3) {$p_3$};
    \node[circle,fill=brown,draw=black] (MP4) at      (4,-3) {$p_4$};

    \path[->,ultra thick,draw=lightgray] (A) edge node[above] {$p_1$} (PA);
    \path[->,ultra thick,draw=purple] (B) edge node[near end,below] {$p_2$} (PB);
    \path[->,ultra thick,draw=red] (A) edge node[right] {$n_1$} (Left);
    \path[->,ultra thick,draw=red] (B) edge node[below,xshift=1mm] {$n_1$} (Right);

    \path[->,ultra thick,draw=pink] (E) edge node[below,near end] {$p_3$} (PE);
    \path[->,ultra thick,draw=brown] (F) edge node[above] {$p_4$} (PF);
    \path[->,ultra thick,draw=orange] (E) edge node[below] {$n_2$} (Left);
    \path[->,ultra thick,draw=orange] (F) edge node[left] {$n_2$} (Right);

    \ifthenelse{\equal{#1}{before}}{
    }{
        \path[->,ultra thick,draw=cyan] (A) edge node[left] {$s_1$} (Pred);
        \path[->,ultra thick,draw=cyan] (B) edge node[above] {$s_1$} (Succ);
    }

    \ifthenelse{\equal{#1}{after2}}{
        \path[->,ultra thick,draw=green] (E) edge node[above] {$s_2$} (Pred);
        \path[->,ultra thick,draw=green] (F) edge node[right] {$s_2$} (Succ);
    }{
    }

    #2
\end{tikzpicture} \end{center}
}

\subsection{Initializing Workspaces}
\label{sec:InitWorkspace}
The \TC{} workspace initially contains only symbols representing input
objects (such as \stringobj{a} in \stringobj{abc}) and information about their
relative position. For example, when comparing the strings \stringobj{ab} and
\stringobj{ef}, the \TC{} workspace is initialized as shown below, where
$x_1$ represents \stringobj{a}, $x_2$ represents \stringobj{b}, $y_1$
represents \stringobj{e}, and $y_2$ represents \stringobj{f}.
\tikzruleall{before}{
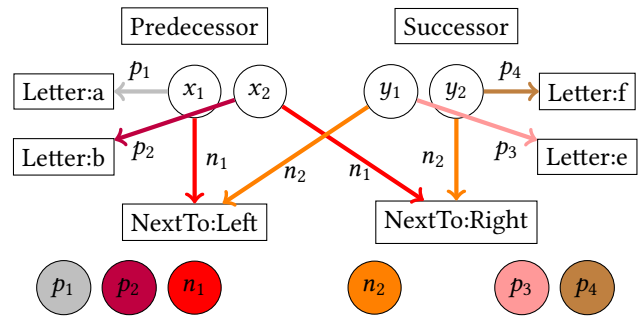
    \node (label) at (0, -4) {\textbf{Triplet Structure T5}};
}
Notably, we also have nodes like \PlatonicLetter{a} representing the Platonic
concept of a particular letter; the fact that $x_1$ is mapped to
\PlatonicLetter{a} corresponds to asserting the unary $\IsA(x_1)$.
We have included nodes for some predicates (specifically Predecessor
and Successor representing the slots of the binary $\LetterSuccessor$
predicate) that have no incoming or outgoing edges. This indicates that, while
\TC{} knows about the concept of predecessor and successor, it has not yet
explicitly recognized any letter-successor pairs in the structure. In the next
section, we will describe how such facts may be inferred from this initial
encoding of the problem via the use of update rules.

\subsection{Modifying the Workspace with Update Rules}
\label{sec:RewriteRules}
\TC{} proceeds to modify the workspace in two ways: (1)~refining the
representation of its inputs, e.g., to infer the
facts $\LetterSuccessor(x_1, x_2)$ and $\LetterSuccessor(y_1, y_2)$
in the above example, and (2)~building
an analogy between its inputs by comparing such inferred facts. Both
types of modifications are implemented using the same framework of update
rules. This section introduces our language for expressing update
rules using a simple example rule of the first kind, deferring discussion of
the second type of modification to \pref{sec:RepresentingAnalogies}. Note that
the language described here for expressing update rules works to define
update rules for any triplet structure. However, we focus our examples
on their use for expressing inference rules for the \TC{} workspace.

Note that, in this section, we will discuss a method of ``hard-coding'' certain
rules to express things like letter-successorship, both because this is how our
current implementation operates, as well as to introduce the notion of update
rules. \pref{sec:Slips} describes more general mechanisms for
making such changes without explicitly enumerating all such rules ahead of
time.

Recall the initial state of the \TC{} workspace for the example of
\stringobj{ab} and \stringobj{ef}, shown above as~\refts{5}.
Consider now the problem of defining a rule that modifies the workspace by
identifying when some letter instance is an alphabetical successor of another.
For example, we may wish to create a rule that marks instances of the letter
`a' and the letter `b' as $\LetterSuccessor$ pairs. In first-order logic, we
might write the desired rule as $\IsA(v_1) \wedge \IsB(v_2) \implies
\LetterSuccessor(v_1, v_2)$.  We have developed a visual domain specific
language~(DSL) for expressing such rules operating on the \TC{} workspace. Our
full DSL is capable of expressing rules containing alternating quantifiers and
other pattern-matching features. We will describe here only a simplified
subset of the language that suffices for the uses in this paper.

Rules in this DSL look like triplet structures themselves (and in fact can be
stored as such), although they are annotated with extra information about which
nodes represent variables to search for and how the structure should be
modified if such variables are found. For example,~\reftsr{1} below shows a
rule which notates \stringobj{a}, \stringobj{b} letter-successor pairs.
\begin{center} \begin{tikzpicture}
    \node[draw=black] (Pred) at (-2,-1) {Predecessor};
    \node[draw=black] (Succ) at (2, -1) {Successor};
    \node[draw=black] (A)    at (-2, 1) {\PlatonicLetter{a}};
    \node[draw=black] (B)    at (2,  1) {\PlatonicLetter{b}};
    \node[circle,dashed,draw=black] (V1) at (-2,0) {$v_1$};
    \node[circle,dashed,draw=black] (V2) at (2,0) {$v_2$};

    \node[circle,dashed,draw=black,fill=red] (VF1) at (-3.5,0) {$vf_1$};
    \node[circle,dashed,draw=black,fill=orange] (VF2) at (3.5,0) {$vf_2$};

    \node[circle,draw=black,fill=teal] (NF2) at (0,0) {$nf_3$};

    \path[->,ultra thick,draw=red] (V1) edge node[left] {$vf_1$} (A);
    \path[->,ultra thick,draw=orange] (V2) edge node[right] {$vf_2$} (B);
    \path[->,ultra thick,draw=teal] (V1) edge node[left] {$nf_3$} (Pred);
    \path[->,ultra thick,draw=teal] (V2) edge node[right] {$nf_3$} (Succ);

    \draw[fill=green,opacity=0.3] (-0.5, -0.5) rectangle (0.5, 0.5);
    \draw[fill=green,opacity=0.3] (-2.75, -0.75) rectangle (-1.75, -0.25);
    \draw[fill=green,opacity=0.3] (2.75, -0.75) rectangle (1.75, -0.25);
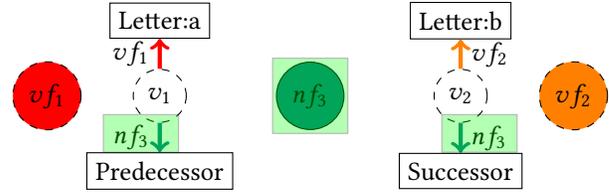
    \node (label) at (0, -1.75) {\textbf{Triplet Structure Rule R1}};
\end{tikzpicture} \end{center}
In such rule diagrams, one first looks at the parts not shaded green. In these
parts, dashed nodes are variables that should be looked for in the structure,
while solid nodes are constants assumed already to exist in the structure. When
this pattern is found in the structure, this is called a \emph{rule match} and
the green nodes and facts can be added. In this case, the rule expresses that
whenever two nodes $v_1$ and $v_2$ are found such that $v_1$ is the letter `a'
and $v_2$ the letter `b', then we can add a new fact node and corresponding
triplet facts which express that $v_2$ is an alphabetical successor of~$v_1$.

For example, we may apply~\reftsr{1} to~\refts{5} by taking the rule assignment
with $v_1 = x_1$, $v_2 = x_2$, $vf_1 = p_1$, and $vf_2 = p_2$.  This
produces a new fact node $nf_3$ which asserts that $x_1$ is a
predecessor of the successor $x_2$. Letting $s_1$ be the generated node
corresponding to $nf_3$, \reftsr{1} transforms~\refts{5} into~\refts{6}
below.
\vspace{3ex}
\tikzruleall{after1}{
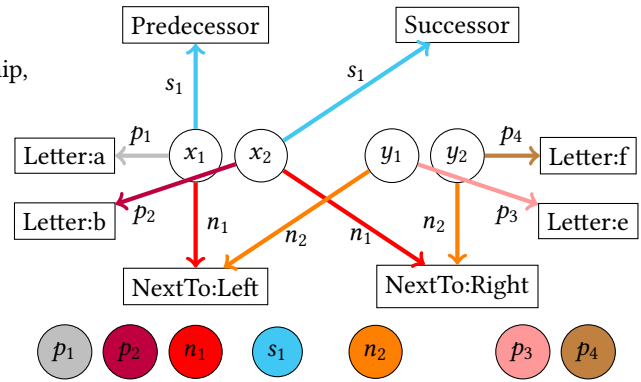
    \node (label) at (0, -4) {\textbf{Triplet Structure T6}};
}

\noindent Similarly,~\reftsr{2} below identifies \stringobj{e}, \stringobj{f} pairs as
successor pairs.
\begin{center} \begin{tikzpicture}
    \node[draw=black] (Pred) at (-2,-1) {Predecessor};
    \node[draw=black] (Succ) at (2, -1) {Successor};
    \node[draw=black] (A)    at (-2, 1) {\PlatonicLetter{e}};
    \node[draw=black] (B)    at (2,  1) {\PlatonicLetter{f}};
    \node[circle,dashed,draw=black] (V1) at (-2,0) {$v_1$};
    \node[circle,dashed,draw=black] (V2) at (2,0) {$v_2$};

    \node[circle,dashed,draw=black,fill=red] (VF1) at (-3.5,0) {$vf_1$};
    \node[circle,dashed,draw=black,fill=orange] (VF2) at (3.5,0) {$vf_2$};

    \node[circle,draw=black,fill=teal] (NF2) at (0,0) {$nf_3$};

    \path[->,ultra thick,draw=red] (V1) edge node[left] {$vf_1$} (A);
    \path[->,ultra thick,draw=orange] (V2) edge node[right] {$vf_2$} (B);
    \path[->,ultra thick,draw=teal] (V1) edge node[left] {$nf_3$} (Pred);
    \path[->,ultra thick,draw=teal] (V2) edge node[right] {$nf_3$} (Succ);

    \draw[fill=green,opacity=0.3] (-0.5, -0.5) rectangle (0.5, 0.5);
    \draw[fill=green,opacity=0.3] (-2.75, -0.75) rectangle (-1.75, -0.25);
    \draw[fill=green,opacity=0.3] (2.75, -0.75) rectangle (1.75, -0.25);
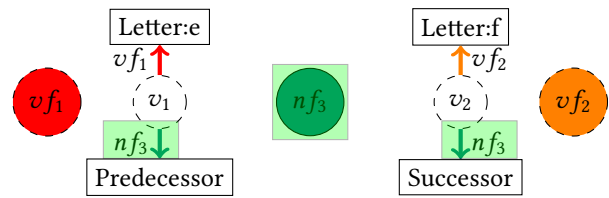
    \node (label) at (0, -1.75) {\textbf{Triplet Structure Rule R2}};
\end{tikzpicture} \end{center}
Applying~\reftsr{2} to~\refts{6} marks $y_1$ and $y_2$ as predecessor and
successor respectively, producing~\refts{7} below.
\tikzruleall{after2}{
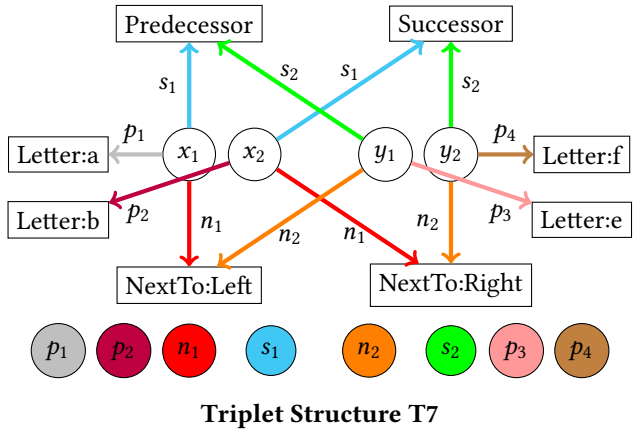
    \node (label) at (0, -4) {\textbf{Triplet Structure T7}};
}

\newcommand{\tikzsimpleabs}[2]{
\begin{center} \begin{tikzpicture}[scale=0.85]
    \node[circle,draw=black] (A) at (-2,0) {$x_1$};
    \node[circle,draw=black] (B) at (-1,0) {$x_2$};
    \node[circle,draw=black] (E) at (1,0) {$y_1$};
    \node[circle,draw=black] (F) at (2,0) {$y_2$};

    \node[circle,draw=black] (Alph1) at (-0.75,-1.25) {$\alpha_1$};
    \ifthenelse{#1>1}{
        \node[circle,draw=black] (Alph2) at (0.75,-1.25) {$\alpha_2$};
    }{}

    \node[draw=black] (Pred) at (-2,1.5) {Predecessor};
    \node[draw=black] (Succ) at (2,1.5)  {Successor};

    \node[draw=black] (Left) at (-2,-2) {$\NextToLeft$};
    \node[draw=black] (Right) at (2,-2) {$\NextToRight$};

    \node[circle,draw=black] (MN1) at (-3,-3) {$n_1$};
    \node[circle,draw=black] (MS1) at (-1.75,-3) {$s_1$};
    \node[circle,draw=black] (MN2) at (-0.5,-3) {$n_2$};
    \node[circle,draw=black] (MS2) at (0.75,-3) {$s_2$};

    \node[circle,fill=purple,draw=black] (MAlphN) at (-3,-5) {$\alpha n$};
    \ifthenelse{#1>2}{
        \node[circle,fill=brown,draw=black] (MAlphS) at (-0.5,-5) {$\alpha s$};
    }{}

    \node[circle,fill=pink,draw=black] (MAlph1) at (2,-3) {$M\alpha_1$};
    \node[circle,fill=cyan,draw=black] (MAlph2) at (3.25,-3) {$M\alpha_2$};

    \node[circle,fill=lightgray,draw=black] (IsAbs) at (4.5,-4) {IsAbs};
    \node[draw=black] (Abs) at (2.5,-5) {Abstraction};

    \path[->,ultra thick,draw=purple] (Alph1) edge
    node[left,xshift=-1mm,yshift=1mm] {$\alpha n$} (Left);
    \ifthenelse{#1>1}{
        \path[->,ultra thick,draw=purple] (Alph2) edge
        node[right,xshift=1mm,yshift=1mm] {$\alpha n$} (Right);
    }{}
    \ifthenelse{#1>2}{
        \path[->,ultra thick,draw=brown] (Alph1) edge
        node[left,yshift=5mm,xshift=-2mm] {$\alpha s$} (Pred);
    }{}
    \ifthenelse{#1=4}{
        \path[->,ultra thick,draw=brown] (Alph2) edge
        node[right,yshift=5mm,xshift=2mm] {$\alpha s$} (Succ);
    }{}

    \path[->,ultra thick,draw=pink] (MN1) edge node[left] {$M\alpha_1$} (MAlphN);
    \path[->,ultra thick,draw=cyan] (MN2) edge
    node[below,xshift=-2mm,fill=white,fill opacity=0.5,text opacity=1] {$M\alpha_2$} (MAlphN);
    \ifthenelse{#1>2}{
        \path[->,ultra thick,draw=pink] (MS1) edge
        node[below,xshift=4mm,yshift=2mm,fill=white,fill opacity=0.5,text opacity=1] {$M\alpha_1$} (MAlphS);
        \path[->,ultra thick,draw=cyan] (MS2) edge node[right] {$M\alpha_2$} (MAlphS);
    }{}

    \path[->,ultra thick,draw=pink] (A) edge node[left] {$M\alpha_1$} (Alph1);
    \path[->,ultra thick,draw=cyan] (E) edge
    node[below,yshift=-1mm,fill=white,fill opacity=0.5,text opacity=1] {$M\alpha_2$} (Alph1);
    \ifthenelse{#1>1}{
        \path[->,ultra thick,draw=pink] (B) edge node[above] {$M\alpha_1$} (Alph2);
        \path[->,ultra thick,draw=cyan] (F) edge node[right] {$M\alpha_2$} (Alph2);
    }{}

    \path[->,ultra thick,draw=lightgray] (MAlph1) edge
    node[left,xshift=3mm,fill=white,fill opacity=0.5,text opacity=1] {IsAbs} (Abs);
    \path[->,ultra thick,draw=lightgray] (MAlph2) edge
    node[right,xshift=-3mm,fill=white,fill opacity=0.5,text opacity=1] {IsAbs} (Abs);

    #2
\end{tikzpicture} \end{center}
}

\subsection{Representing Analogies in Triplet Structures}
\label{sec:RepresentingAnalogies}

This section discusses how $\TC$ represents and makes analogies in its
triplet-structure workspace. Analogies are represented
as
\emph{abstractions}, similar to the abstract letter strings
in~\pref{sec:WhatIsAnalogyMaking}. Intuitively, instructed to form an analogy
between \stringobj{abc} and \stringobj{efg}, \TC{} forms a shared abstract
representation of them both, roughly of the form \stringobj{(?1)(?2)(?3)} with
additional information such as \stringobj{(?2)} is a successor of
\stringobj{(?1)}. It then adds facts stating that, for example, both the original
\stringobj{a} and \stringobj{e} are \emph{instances of} this more abstract
\stringobj{(?1)} object. Two input objects
\emph{correspond} if they are instances of the same abstract node.

Consider the example from~\pref{sec:RewriteRules}, where we are forming
an analogy between two letter strings \stringobj{ab} and \stringobj{ef}.
Suppose the current \TC{} workspace is represented by~\refts{8} below, which is
identical to~\refts{7} except with a few of the nodes/relations removed for
ease of exposition.  \begin{center} \begin{tikzpicture}[scale=1]
    \node[circle,draw=black] (A) at (-2,0) {$x_1$};
    \node[circle,draw=black] (B) at (-1,0) {$x_2$};
    \node[circle,draw=black] (E) at (1,0) {$y_1$};
    \node[circle,draw=black] (F) at (2,0) {$y_2$};

    \node[draw=black] (Pred) at (-2,1) {Predecessor};
    \node[draw=black] (Succ) at (2,1)  {Successor};

    \node[draw=black] (Left) at (-2,-1) {$\NextToLeft$};
    \node[draw=black] (Right) at (2,-1) {$\NextToRight$};

    \node[circle,fill=red,draw=black] (MN1) at (-2,-2) {$n_1$};
    \node[circle,fill=cyan,draw=black] (MS1) at (-0.75,-2) {$s_1$};
    \node[circle,fill=orange,draw=black] (MN2) at (0.75,-2) {$n_2$};
    \node[circle,fill=green,draw=black] (MS2) at (2,-2) {$s_2$};

    \path[->,ultra thick,draw=red] (A) edge node[left] {$n_1$} (Left);
    \path[->,ultra thick,draw=red] (B) edge node[below,xshift=3mm] {$n_1$} (Right);
    \path[->,ultra thick,draw=cyan] (A) edge node[left] {$s_1$} (Pred);
    \path[->,ultra thick,draw=cyan] (B) edge node[above,xshift=3mm,yshift=0.5mm] {$s_1$} (Succ);

    \path[->,ultra thick,draw=orange] (E) edge node[below,xshift=-2mm] {$n_2$} (Left);
    \path[->,ultra thick,draw=orange] (F) edge node[right] {$n_2$} (Right);
    \path[->,ultra thick,draw=green] (E) edge node[above,xshift=-2mm,yshift=0.5mm] {$s_2$} (Pred);
    \path[->,ultra thick,draw=green] (F) edge node[right] {$s_2$} (Succ);
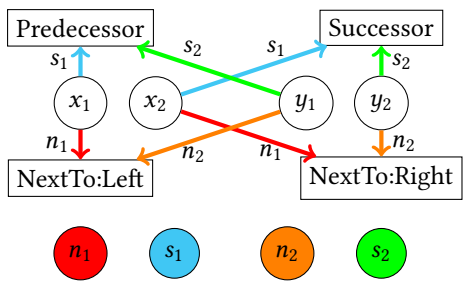
    \node (label) at (0, -3) {\textbf{Triplet Structure T8}};
\end{tikzpicture} \end{center}

An \emph{analogy} between the two strings \stringobj{ab} and \stringobj{ef}
might determine that $x_1$ corresponds to $y_1$ and $x_2$ to $y_2$, because they
both form instances of a more abstract type of ``two successive letters next to
each other.'' This is represented in the \TC{} workspace as~\refts{9}
below. For clarity, we have only shown the newly-added facts (i.e.\
those that make up the abstraction), although the facts from~\refts{8} would
still be present.
\tikzsimpleabs{4}{
    \node (label) at (0, -6) {\textbf{Triplet Structure T9}};
}
In~\refts{9}, we have added new nodes $\alpha_1$ and $\alpha_2$ to represent
the abstract type of which $x_1, y_1$ and $x_2, y_2$ respectively are instances
of. We have also abstracted the fact nodes that correspond to each other into
nodes $\alpha n$ and $\alpha s$. These abstract fact nodes each express the
same fact about the abstract $\alpha_1$ and $\alpha_2$ as the original, or
\emph{concrete}, fact nodes expressed about, e.g., $x_1$ and $x_2$.  Finally,
we have added fact nodes $M\alpha_1$ and $M\alpha_2$ that map the concrete
nodes in each instance to their abstract counterparts. For bookkeeping reasons
in the structure, we label each of these as $Abstraction$s so we can keep track
of which nodes in the workspace are abstract vs.\ provided in the input.

From~\refts{9} above, we can extract the analogy that $x_1$ corresponds to
$y_1$ because both are instances of the abstract $\alpha_1$ node, and similarly
for $x_2$, $y_2$, and $\alpha_2$.

\newcommand{\tikzmapperfirstblock}[2][noslip]{
    \node[circle,dashed,draw=black] (A) at (-1.5,0) {$A$};
    \node[circle,dashed,draw=black] (B) at (1.5,0) {$B$};
    \node[circle,#2,draw=black] (AlphaAB) at (0,1) {$\alpha AB$};
    \ifthenelse{\equal{#1}{noslip}}{
        \node[circle,dashed,draw=black] (C) at (0,3) {$C$};
    }{
        \node[circle,dashed,draw=black] (C1) at (-1.5,4) {$C_1$};
        \node[circle,dashed,draw=black] (C2) at (1.5,4) {$C_2$};
        \node[circle,#2,draw=black] (AlphaC) at (0,3) {$\alpha C$};
    }
}

\newcommand{\tikzmappersecondblock}[1]{
    \node[circle,dashed,draw=black,fill=red] (MA) at (3,0) {$M_A$};
    \node[circle,dashed,draw=black,fill=orange] (MB) at (4,0) {$M_B$};
    \node[circle,#1,draw=black,fill=cyan] (AlphaMAB) at (3.5, 2.5) {$\alpha M_{AB}$};
}

\newcommand{\tikzmapperthirdblock}[1]{
    \node[circle,#1,draw=black,fill=yellow] (MAlphaA) at (5.5,1) {$M\alpha A$};
    \node[circle,#1,draw=black,fill=purple] (MAlphaB) at (7,1) {$M\alpha B$};

    \node[draw=black] (Abstraction) at (6.25,3) {Abstraction};

    \node[circle,#1,draw=black,fill=lightgray] (IsAbs) at (6.25,0) {IsAbs};
}

\newcommand{\tikzmapperedges}[1][noslip]{
    \ifthenelse{\equal{#1}{noslip}}{
        \path[->,ultra thick,draw=red] (A) edge node[left] {$M_A$} (C);
        \path[->,ultra thick,draw=orange] (B) edge node[right] {$M_B$} (C);
        \path[->,ultra thick,draw=yellow] (A) edge
        node[below,xshift=1mm] {$M\alpha A$} (AlphaAB);
        \path[->,ultra thick,draw=purple] (B) edge
        node[below,xshift=-1mm] {$M\alpha B$} (AlphaAB);
        \path[->,ultra thick,draw=cyan] (AlphaAB) edge node[below] {$\alpha M_{AB}$} (C);
    }{
        \path[->,ultra thick,draw=red] (A) edge node[left] {$M_A$} (C1);
        \path[->,ultra thick,draw=orange] (B) edge node[right] {$M_B$} (C2);
        \path[->,ultra thick,draw=yellow] (A) edge
        node[above,yshift=1mm,xshift=-1mm] {$M\alpha A$} (AlphaAB);
        \path[->,ultra thick,draw=yellow] (C1) edge
        node[above,xshift=2mm,yshift=-1mm] {$M\alpha A$} (C);
        \path[->,ultra thick,draw=purple] (B) edge
        node[above,yshift=1mm,xshift=1mm] {$M\alpha B$} (AlphaAB);
        \path[->,ultra thick,draw=purple] (C2) edge
        node[above,yshift=-1mm,xshift=-2mm] {$M\alpha B$} (C);
        \path[->,ultra thick,draw=cyan] (AlphaAB) edge node[right] {$\alpha M_{AB}$} (C);
    }
    \path[->,ultra thick,draw=yellow] (MA) edge
    node[left,xshift=3mm,fill=white,fill opacity=0.5,text opacity=1] {$M\alpha A$} (AlphaMAB);
    \path[->,ultra thick,draw=purple] (MB) edge
    node[right,xshift=-3mm,fill=white,fill opacity=0.7,text opacity=1] {$M\alpha B$} (AlphaMAB);
    \path[->,ultra thick,draw=lightgray] (MAlphaA) edge
    node[left,xshift=3mm,fill=white,fill opacity=0.5,text opacity=1] {IsAbs} (Abstraction);
    \path[->,ultra thick,draw=lightgray] (MAlphaB) edge
    node[right,xshift=-3mm,fill=white,fill opacity=0.5,text opacity=1] {IsAbs} (Abstraction);
}

\subsection{Rules for Making Analogies in a Triplet Structure}
\label{sec:ImplementationAnalogyRules}
We now turn our
attention to designing rules for forming such analogies. All such rules will be
of the form discussed in \pref{sec:RewriteRules}. Each rule application makes a
small change to the structure; for example, abstracting two concrete nodes
together, or lifting a single concrete fact to the abstraction. These rules
create a search space that can be explored using heuristics.

We have found that all of the rules necessary for abstraction-forming can be
formed as variations on the following \emph{Begin Analogy rule},~\reftsr{3},
which starts a new abstraction.
\begin{center} \begin{tikzpicture}[scale=0.9]
    \tikzmapperfirstblock{}
    \tikzmappersecondblock{}
    \tikzmapperthirdblock{}
    \tikzmapperedges{}

    \draw[fill=green,opacity=0.3] (-1, 0.4) -- (1, 0.4) -- (0, 2.5) -- cycle;
    \draw[fill=green,opacity=0.3] (2.4, 0.5) rectangle (4.6, 3.3);
    \draw[fill=green,opacity=0.3] (4.9, -0.7) rectangle (7.6, 2.75);

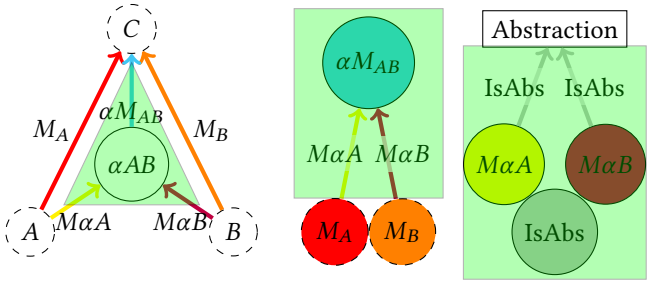
    \node (label) at (2, -1.25) {\textbf{Triplet Structure Rule R3}};
\end{tikzpicture} \end{center}
Here, the variable nodes $A$ and $B$ represent the concrete nodes that should
correspond to each other in the analogy, like $x_1$ and $y_1$ in the previous
example. The variable node $C$ is the field that they both share. For example,
in the previous example $C$ might be $\NextToLeft$, because both $x_1$ and $y_1$
are mapped to $\NextToLeft$ by fact nodes $n_1$ and $n_2$, which in
turn map to $M_A$ and $M_B$, respectively, in the rule above.

Applying~\reftsr{3} to~\refts{9} may produce the start of an analogy shown
below in~\refts{10}. For clarity, we have left out the facts between
concrete nodes.
\tikzsimpleabs{1}{
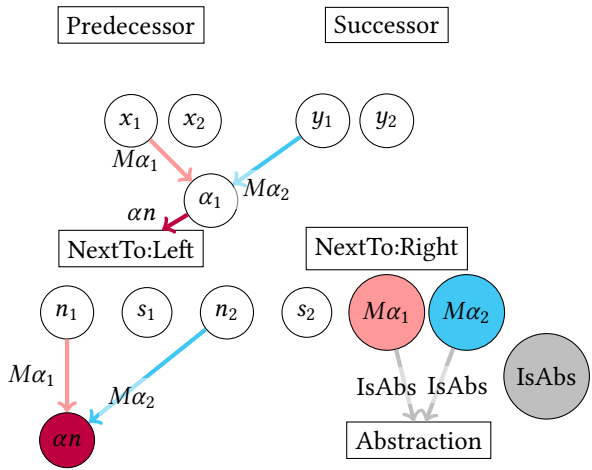
    \node (label) at (0, -6) {\textbf{Triplet Structure T10}};
}

By shading different subsets of the nodes green, we can modify~\reftsr{3} into
a variety of rules for \emph{extending analogies.} For example,~\reftsr{4}
below ``follows'' a fact node from an existing analogy to map two new concrete
nodes to each other.
\begin{center} \begin{tikzpicture}[scale=0.9]
    \tikzmapperfirstblock{}
    \tikzmappersecondblock{dashed}
    \tikzmapperthirdblock{dashed}
    \tikzmapperedges{}

    \draw[fill=green,opacity=0.3] (-1, 0.4) -- (1, 0.4) -- (0, 2.5) -- cycle;

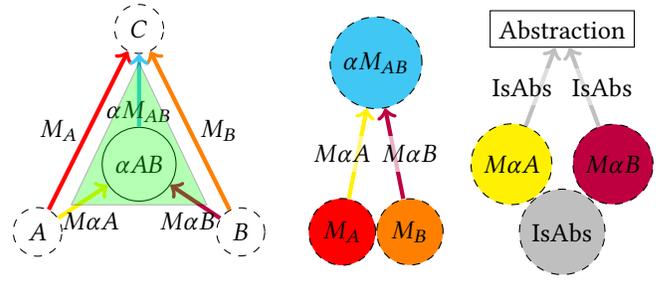
    \node (label) at (2, -1.25) {\textbf{Triplet Structure Rule R4}};
\end{tikzpicture} \end{center}

Applying~\reftsr{4} to our running structure with $A = x_2$, $B = y_2$, and $C
= \NextToRight$ would extend the analogy to include $x_2$ and $y_2$ by
``following'' the $\NextToRight$ relation, producing~\refts{11} below.
\tikzsimpleabs{2}{
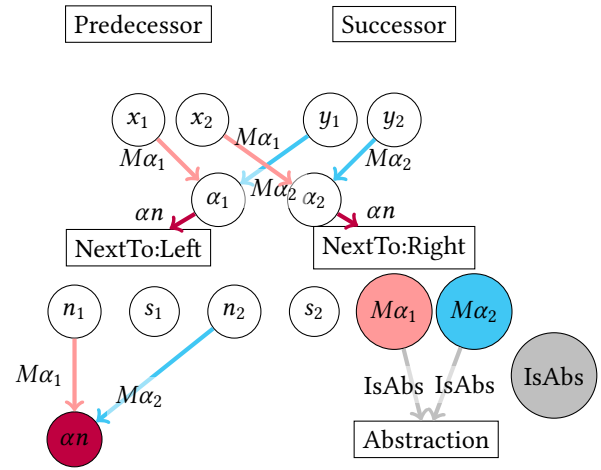
    \node (label) at (0, -6) {\textbf{Triplet Structure T11}};
}

\noindent Similarly, by shading just the $\alpha M_{AB}$ node, we get~\reftsr{5} that
adds a new fact node to the abstraction.
\begin{center} \begin{tikzpicture}[scale=0.9]
    \tikzmapperfirstblock{dashed}
    \tikzmappersecondblock{}
    \tikzmapperthirdblock{dashed}
    \tikzmapperedges{}

    \draw[fill=green,opacity=0.3] (-0.5, 1.5) -- (0.5, 1.5) -- (0, 2.5) -- cycle;
    \draw[fill=green,opacity=0.3] (2.4, 0.5) rectangle (4.6, 3.3);

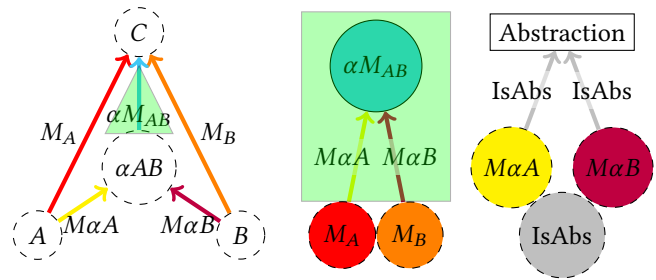
    \node (label) at (2, -1.25) {\textbf{Triplet Structure Rule R5}};
\end{tikzpicture} \end{center}
Applying~\reftsr{5} to the previous abstraction with $A = x_1$, $B = y_1$, $\alpha AB =
\alpha_1$, $C = Predecessor$, $M_A = s_1$, and $M_B = s_2$ allows us to
associate $s_1$ and $s_2$ with each other, producing~\refts{12} as
shown below.
\tikzsimpleabs{3}{
    \node (label) at (0, -6) {\textbf{Triplet Structure T12}};
}

Finally, by shading just the blue fact edge from $\alpha AB$ to $C$, we
get~\reftsr{6} that lifts a single fact into the abstraction.
\begin{center} \begin{tikzpicture}[scale=0.9]
    \tikzmapperfirstblock{dashed}
    \tikzmappersecondblock{dashed}
    \tikzmapperthirdblock{dashed}
    \tikzmapperedges{}

    \draw[fill=green,opacity=0.3] (-0.5, 1.5) -- (0.5, 1.5) -- (0, 2.5) -- cycle;

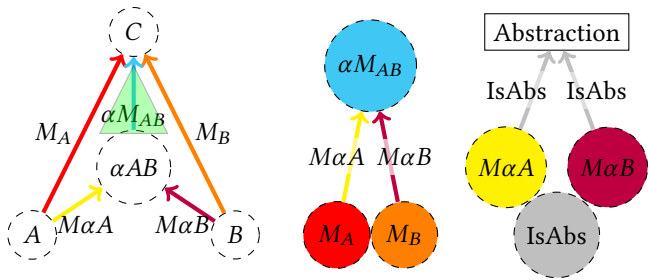
    \node (label) at (2, -1.25) {\textbf{Triplet Structure Rule R6}};
\end{tikzpicture} \end{center}

Applying this rule with $\alpha AB = \alpha_2$, $C = Successor$, and $\alpha
M_{AB} = \alpha s$ completes the abstraction, giving the final abstraction
we saw earlier in~\refts{9}, reproduced below.
\vspace{2ex}
\tikzsimpleabs{4}{
    \node (label) at (0, -6) {\textbf{Triplet Structure T9}};
}

We encourage the motivated reader to consider interpretations of different
shadings. Of particular interest is if a node like $B$ is shaded green,
which corresponds to \emph{completing} an analogy: constructing a concrete node
that plays a particular role in an existing analogy. This shading of
the rule prototype is how $\TC$ generated the code in the
bottom right of~\pref{fig:GEMM} and~\pref{fig:API}.

\subsection{Higher-Order Analogies and Slips}
\label{sec:Slips}
We described in~\pref{sec:RewriteRules} that rules could be used to infer new
facts, such as $\LetterSuccessor(x_1, x_2)$. Such facts could later be used in
analogies, e.g., to compare \stringobj{ab} and \stringobj{ef} as ``two-letter
strings where the letters satisfy the $\LetterSuccessor$ relation''. To do
this, we had to first explicitly add the facts $\LetterSuccessor(x_1, x_2)$ and
$\LetterSuccessor(y_1, y_2)$. In general, this approach requires us to
explicitly enumerate rules for all such relations used in our
analogies. This section considers a more general approach
based on forming analogies between types in the structure.

At first glance, it is tempting to resolve the issue using a general
transitivity rule such that, for example, if $x_1$ is an instance of $T_1$,
$x_2$ is an instance of $T_2$, and there is some fact $R(T_1, T_2)$, then we
can add $R(x_1, x_2)$ as well. For example, if
$\LetterSuccessor(\PlatonicLetter{a}, \PlatonicLetter{b})$ and $x_1$ was an
instance of $\PlatonicLetter{a}$, $x_2$ an instance of $\PlatonicLetter{b}$,
then the rule would infer $\LetterSuccessor(x_1, x_2)$ as desired.

However, facts about \emph{types} may not be valid or well-defined when applied
to \emph{instances of those types.} For example, when forming an analogy
involving both numerical value and color we may have two nodes be instances of
opposite numbers, e.g., $-1$ and $1$, or opposite colors, e.g., black and
white. If we were to directly use transitivity to say that the two nodes were
simply ``opposites,'' we would lose important information because we would not
know whether they were opposite \emph{numbers} or \emph{colors.} Similarly,
consider forming an analogy between the pairs \stringpair{abc}{cba} and
\stringpair{efg}{gfe}. Fundamentally, what we want to express is that the
letters in the first string satisfy \emph{either} $\LetterSuccessor$ \emph{or}
$\LetterPredecessor$, and that those in the second string satisfy \emph{the
opposite.}

To express such scenarios naturally, we need a way to \emph{include the types
in the analogy,} i.e., make a \emph{type slip.} In the mapping rules
shown so far, we require the both concrete nodes $A$ and $B$ to be of the same
type $C$. However, we can define new mapping rules, using the template of~\reftsr{7},
that allow the type itself to be abstracted and become part of the analogy.
In the example discussed, we could have $\LetterSuccessor$ be $C_1$ and
$\LetterPredecessor$ be $C_2$.
\begin{center} \begin{tikzpicture}[scale=0.8]
    \tikzmapperfirstblock[slip]{}
    \tikzmappersecondblock{}
    \tikzmapperthirdblock{}
    \tikzmapperedges[slip]

    \draw[fill=green,opacity=0.3] (-1.25, 0.3) rectangle (1.25, 3.8);
    \draw[fill=green,opacity=0.3] (2.4, 0.5) rectangle (4.6, 3.25);
    \draw[fill=green,opacity=0.3] (4.8, -0.7) rectangle (7.7, 2.75);

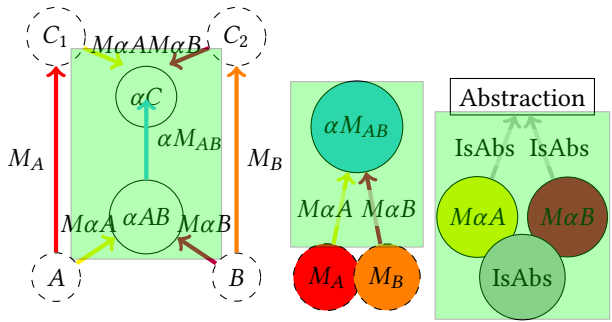
    \node (label) at (2, -1.25) {\textbf{Triplet Structure Rule R7}};
\end{tikzpicture} \end{center}

Although allowing for creative analogies, such an approach significantly
increases the search space. To control this, one can make \emph{compound
analogies,} using type slips only for small sub-analogies with smaller search
spaces. Then those analogies are used to define types with which to build up
larger ones.  For example, we could use a type slip to learn the abstract type
of ``pairs of nodes which are instances of types that have $Successor$
relation,'' i.e., effectively re-learn the $\LetterSuccessor$ relation on its
own. Analogy-making rules could then be used to note that both $x_1, x_2$ and
$y_1, y_2$ form an instance of this abstract type, and then the abstract type
can be used exactly like Predecessor and Successor in future analogies.

\subsection{Prototype Implementation and Optimizations}
We have implemented a proof-of-concept version of \TC{} in Python, with
hotspots written in C++. We have run the demonstrations discussed in
\pref{sec:WhatIf} on our prototype to verify that such analogies can be
found, represented, and completed by \TC{}.

The project is divided into distinct components, including: \texttt{TSLib},
a library for declaring triplet structures and rules operating on them;
\texttt{Abstracter}, a collection of such rules which can be applied to build
up analogies; and \texttt{TSRuntime}, an interface for efficiently applying
rules to a triplet-structure workspace and can optimize pattern matching,
e.g., by only checking parts of the structure that have changed since the
pattern was last checked against.

In its full generality, our update rule DSL is Turing-complete, and capable
of expressing rules matching complex patterns.  One useful feature of
our system is its ability to express \emph{consistency
constraints} or desired invariants on the structure. For example, we may
want to ensure that any symbol can \emph{either} be an instance of
\PlatonicLetter{a} \emph{or} \PlatonicLetter{b}, but never both at the same
time. \TC{} supports \emph{consistency rules}, which are just like normal
update rules except (1)~they are checked for matches \emph{every time} the
structure is modified, and (2)~they force $\TC$ to backtrack when they
match, i.e.\ undo the last modification.

We optimized our implementation for efficient search through possible
applications of update rules, described in more detail
in~\pref{app:Implementation}. The most impactful optimization has been the
use of \emph{differential matching,} where the design of our update rules
allows us to restrict our search to only those assignments that make use
of the facts added since we last checked for assignments, significantly
reducing the amount of redundant time spent searching. More optimizations
are possible in the future, both in speeding up such a tree-search approach
to analogy-making as well as by investigating other architectures for
making analogies within \TC{} (see~\pref{sec:HeuristicsForSWA}).

\section{Efficient Software Analogies with \TC{}}
This section discusses practical considerations with the
application of \TC{} to make analogies of the form shown in~\pref{sec:WhatIf}.
We focus on three particular factors, (i) the representation of source
code as a triplet structure, (ii) the use of other sources of reasoning and
information, and (iii) heuristics for finding analogies. In each section,
we begin with a description of our current solution, then discuss our vision of
what a future implementation may be able to accomplish.

\subsection{Representing Source Code as Triplet Structures}
\label{sec:RepresentingCode}
Currently, given a source file we perform a lightweight lexical analysis before
encoding it in the structure. If more information about the meaning of some of
the resulting lexemes is known, we can include that as well. For example, given
a source file consisting only of the statement \lstinline{name=user.name}, we
might encode it as shown in~\refts{12}.
\begin{center}
\begin{tikzpicture}[scale=0.95]
    \node[draw=black] (Left) at (1,2) {$\NextToLeft$};
    \node[draw=black] (Right) at (3.5,2) {$\NextToRight$};

    \node[draw=black] (isname) at (-3.25,2) {Is``name''};
    \node[draw=black] (iseq) at (-1.75,2) {Is``=''};
    \node[draw=black] (isuser) at (-3.25,1) {Is``user''};
    \node[draw=black] (isdot) at (-1.75,1) {Is``.''};

    \node[draw=black] (File) at (-2.75,-1.25) {$File$};
    \node[draw=black] (Member) at  (-1,-1.25) {$FileMember$};

    \node[draw=black] (Object) at   (2,-1.25) {$Object$};
    \node[draw=black] (Access) at (3.4,-1.25) {$Access$};
    \node[draw=black] (Field) at (4.75,-1.25) {$Field$};

    \node[draw=black] (fileext) at (-2,0) {$file.ext:$};

    \node[draw=black] (namevar) at (0,0) {$name$};
    \node[draw=black] (eq) at (1,0) {$=$};
    \node[draw=black] (user) at (2,0) {$user$};
    \node[draw=black] (dot) at (3,0) {$.$};
    \node[draw=black] (name) at (4,0) {$name$};

    \sclr{1}{black}{namevar}{eq}
    \sclr{2}{red}{eq}{user}
    \sclr{3}{green}{user}{dot}
    \sclr{4}{orange}{dot}{name}

    \scedge{5}{cyan}{user}{Object}{right}
    \scedge{5}{cyan}{dot}{Access}{right}
    \scedge{5}{cyan}{name}{Field}{right}

    \scedge{6}{brown}{fileext}{File}{left}
    \scedge{6}{brown}{namevar}{Member}{left}
    \scedge{6}{brown}{eq}{Member}{left}
    \scedge{6}{brown}{user}{Member}{left}
    \scedge{6}{brown}{dot}{Member}{left}
    \scedge{6}{brown}{name}{Member}{left}

    \scedge{7}{gray}{namevar}{isname}{left,near end}
    \scedge{8}{yellow}{eq}{iseq}{right,near end,yshift=2mm,xshift=-2mm}
    \scedge{9}{magenta}{user}{isuser}{left,near end}
    \scedge{10}{purple}{dot}{isdot}{left,near end}
    \scedge{11}{pink}{name}{isname}{left,near end}

    \node (label) at (0.5, -2) {\textbf{Triplet Structure T12}};
\end{tikzpicture}
\end{center}
For brevity, 11 fact nodes are not explicitly shown. However, their
existence is implied by the colors and labels on the edges. We first create a
node in the structure representing the file. For each lexeme in the file we add
a corresponding node. Each lexeme node is marked as a member of the
corresponding file, and their relative positions are specified using
$\NextToLeft$ and $\NextToRight$. There are four nodes representing Platonic
strings (or `tokens'), which play the same role as the \PlatonicLetter{a} nodes
in~\pref{sec:RewriteRules} or a unary $\IsA(x)$ predicate. In this example, we
assume additional information about the language,
namely that \lstinline{user.name} represents an access of the \lstinline{name}
field of the \lstinline{user} object.

Notably, such a lexical analysis can usually be developed quite quickly even
for new programming languages. At its simplest, it can be implemented as just
splitting the source file based on whitespace and special characters such as
\lstinline{*}. This allows for \TC{} to be applicable to nascent DSLs and other
languages where a full compiler and AST generator has not yet been developed,
or to work with syntactically-invalid programs. As the language tooling grows
in maturity or parts of the programs become syntactically valid, more detailed
information can be produced from the lexing pass and included in the structure,
such as the object-field access notated in the above example.

\paragraph{Future Work: Full ASTs} While the progressive-lexing style of
encoding strikes a nice balance between flexibility and richness, if a full AST
is readily available for the code in question, then this can be used to produce
a richer encoding of the structure. \refts{13} shows how an AST for the
\lstinline{name=user.name} example might be encoded as a triplet structure.
\begin{center}
\begin{tikzpicture}[scale=0.95]
    \node[draw=black] (AssignTo) at  (-3, 1.25) {$AssignTo$};
    \node[draw=black] (Assignment) at (-1, 1.25) {$Assignment$};
    \node[draw=black] (AssignFrom) at (1, 1.25) {$AssignFrom$};

    \node[draw=black] (IdentifierName) at  (3,-2) {$Identifier``name''$};
    \node[draw=black] (IdentifierUser) at (-2,-2) {$Identifier``user''$};

    \node[draw=black] (MemberExpr) at  (3.5, 1) {$MemberExpr$};
    \node[draw=black] (Object) at      (2.5, -0.25) {$Object$};
    \node[draw=black] (Property) at      (4, -0.25) {$Property$};

    \node[draw=black] (eq) at (0,0.25) {$=$};
    \node[draw=black] (namevar) at (-0.75,-0.25) {$name$};
    \node[draw=black] (dot) at (1,-0.25) {$.$};
    \node[draw=black] (user) at (0,-1) {$user$};
    \node[draw=black] (name) at (2,-1) {$name$};

    \scedge{1}{gray}{eq}{Assignment}{right,near end}
    \scedge{1}{gray}{namevar}{AssignTo}{left,near end}
    \scedge{1}{gray}{dot}{AssignFrom}{right,near end}

    \scedge{2}{cyan}{dot}{MemberExpr}{near end}
    \scedge{2}{cyan}{user}{Object}{left,near end}
    \scedge{2}{cyan}{name}{Property}{right,near end}

    \scedge{3}{red}{namevar}{IdentifierName}{left,near end}
    \scedge{4}{green}{name}{IdentifierName}{right,near end}
    \scedge{5}{yellow}{user}{IdentifierUser}{left,near end}

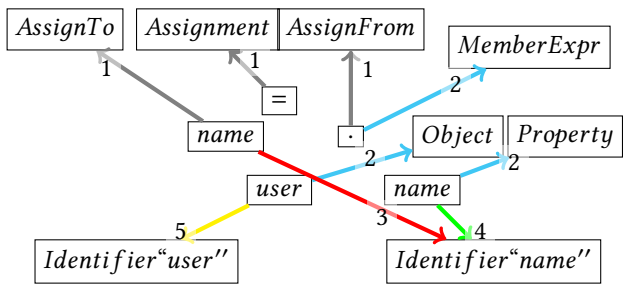
    \node (label) at (0.5, -3) {\textbf{Triplet Structure T13}};
\end{tikzpicture}
\end{center}

\paragraph{Future Work: Multiple Granularities} One particularly exciting area
of future work for our encoding is to allow the granularity of the encoding to
change dynamically, as the analogy process is proceeding. In this model, the
workspace would initially begin with only a listing of the names of files and
folders in the root directory of the project(s). As analogy-making proceeds,
the contents of files may be added to the structure either randomly or
according to activity from the analogy-making process itself. For example, if two
files are named the same, they may be mapped together indicating their
importance to the analogy and hinting to the system that the file contents
might be important as well. If more semantic information about the programming
languages is known, this type of multi-granularity encoding can be used at that
level as well. For example, a file could be loaded first as just a list of
functions contained in it. If two functions seem similar based on
their signatures, then we expand them and include the full associated code in
the workspace.

Our current implementation supports such real-time modifications to
the structure, like adding a new file's contents halfway during an analogy run.
However, the heuristics for knowing when to do such are not yet developed, so we just
add the full contents of the files to it at the start (limiting us to small-ish
projects).

\subsection{Use of Other Engines}
Many other reasoning engines for both natural language and software source-code
exist, including logic-based techniques (such as Cyc~\cite{cyc}), abstract
interpretation~\cite{DBLP:conf/popl/CousotC77}, and statistical techniques
(such as deep learning~\cite{Goodfellow:DeepLearning2016}). We designed \TC{}
with the specific goal of easily integrating the knowledge stored in
such tools with the analogy-making process. In particular, while we focused
in~\pref{sec:RewriteRules} on the usage of update rules to make inferences
about relations such as $\LetterSuccessor$, there is no requirement
that modifications to the structure come from such a update rule. Instead,
other reasoning engines can provide their own insights into the problem at
hand, which can then be translated into triplet facts and added to the
structure. Such added facts are used in analogies just like any other facts.

For example, a statistical model might be used to identify when words used are
synonyms, which can be encoded into the workspace, e.g., using a Synonym
predicate or by stating that both words are instances of some common semantic
notion. Similarly, a logical inference engine like Cyc~\cite{cyc} might be used
to understand comments in, e.g., function docstrings to identify the semantic
meaning behind different functions, or infer the effects of applying multiple
different functions in sequence. This information can be encoded back into the
triplet structure and used to make analogies.

\paragraph{Future Work: A Unified Workspace} We envision that
the triplet structure workspace of \TC{} can serve as a shared workspace among
a host of reasoning engines.  The core analogy-making rules, which operate
structurally, and are in some sense oblivious to the semantics of the
relations, can act as glue that can help synthesize inferences across
distinct reasoning engines. Reasoning engines can operate independently,
inferring new facts and adding them to the structure.  Furthermore, they may be
able to use the facts inferred by other engines or even predicted to be true
via analogy to further their own reasoning, resulting in a virtuous cycle of
cooperation between different engines.

\paragraph{Future Work: Probabilistic Triplet Structures}
To better interface with statistical models, it may be desirable to associate
with each triplet fact a corresponding real-valued probability representing
\TC{}'s confidence in that inferred fact. As future inferences are made, their
confidence values can be computed as a function of the confidence in that
specific step as well as the confidence in the facts it relies on to make that
inference. However, as argued by the Cyc authors~\cite{cyc}, one should be
careful in treating such numbers as a measure of truthfulness of a claim, and
instead only as a representation of one's epistemological uncertainty about
immediate observations of the environment.

\subsection{Heuristics for and Identification of Strong Software Analogies}
\label{sec:HeuristicsForSWA}
To guide the search process, we need a notion of the strength of an analogy.
One initial approach to this is to define a stronger analogy as one with more
shared facts. This idea can be improved by weighting different types of
relations with an importance. For a somewhat extreme example, we may prefer to
map two functions together that share a relation of ``called by analogous
methods'' rather than simply ``share the same first letter.'' This approach is
exemplified by the notion of a \emph{Slipnet} in the Copycat architecture,
which assign numerical, \emph{a priori} importances to each type of relation.

For scenarios where the goal of the analogy is to generate some
\emph{completion} (as in~\pref{sec:WhatIfGEMM}~and~\ref{sec:APIMigration}), we
have found that an even stronger heuristic measure of analogy depth is to
check if a full completion to the original problem can be made from it.
In our experience, when we had bugs with our search process, we found that the
completions produced in~\pref{fig:GEMM}~and~\ref{fig:API} would either (i)
contain very few nodes, or (ii) contain many nodes that the system thought
would be there, but was not able to actually infer (consistently) what tokens
they should be. This can be used to indicate when an analogy may fit the two
examples very well, but does not generalize to the prompt.

\paragraph{Future Work: Cognitive Models} In the long term, we are excited
about the possibility of having multiple workers that operate concurrently,
either working on separate attempts at making such analogies, finding
inconsistencies in analogies made by others, or working together to produce a
strong analogy.  Such a system may take inspiration from psychology-inspired
architectures and theories, like LIDA~\cite{LIDA} and Global Workspace
Theory~\cite{GWT}. Such a system could be supplemented via the use of something
akin to abduction, where rules can request the help of other rules. For instance, an
abstraction rule expressing that it would be able to abstract two things
together if only one of them were also a letter successor would
encourage rules to search for letter successor facts relating to that node.
This abductive inference is similar to the execution of the Slipnet and
Coderack in the original Copycat algorithm.

\paragraph{Future Work: Meta-Reasoning} \TC{} may be extended to support
meta-reasoning similar to that of Metacat~\cite{marshall2000metacat}, where the
system learns to recognize common `snags' that it can store and refer back to
when it encounters a new problem that is challenging in the same way. It may
recognize such ``challenging in the same way'' instances via a sort of
meta-\TC{}, forming analogies between solver states. In such a way, it would be
able to effectively introspect on its own solving process via a (copy of) the
solver itself.  Such a system may be aided by the fact that our \TC{} update
rules can be expressed \emph{within triplet structures themselves.}

\section{Related Work}
\label{sec:RelatedWork}

\noindent\textbf{Analogy making}
Analogies, and the more general class of
\emph{metaphors}~\cite{lakoff2008metaphors,lakoff2000mathematics}, have been
studied extensively in cognitive science.
The primary inspiration for this paper was the Copycat algorithm of Melanie
Mitchell and Douglas
Hofstadter~\cite{hofstadter1995fluid,hofstadter1994copycat}, which we
generalized into the \TC{} analogy-making algorithm presented here. Mitchell's
original source code~\cite{mitchell-copycat} is written in a now-defunct
dialect of Common Lisp for which we could not obtain an interpreter.
Thankfully, there are a number of more-modern
ports~\cite{farg-copycat,ajhager-copycat} which we were able to reference and
run. We originally intended to adapt these implementations for use on program
source code, but quickly found that the Copycat algorithm is highly specific
for the letter string domain. Adding support even for upper-case letters, for
example, turned out to be a significant project, touching almost every file in
the implementation. This is because Copycat implements relations like
$\LetterPredecessor$ by special-case checks and data structures throughout the
code, not as the sort of arbitrary relations that are more familiar in systems
based on first-order logic. These issues motivated the construction of our
\TC{} analogy-making algorithm.

Beyond Copycat, there are a number of other analogy-making algorithms explored
in the cognitive science and philosophy literature, such as
SME~\cite{falkenhainer1989structure}, ACME~\cite{holyoak1989analogical},
Winston's Analogy-Maker~\cite{winston1980learning},
and the Geometric Analogy-Maker~\cite{evans1964heuristic}.
Below we will discuss SME, but ACME and Winston's algorithm have a similar
operation. The Geometric Analogy-Maker is interesting and unique, incorporating some
amount of grouping and representational manipulation similar to that of
Copycat, although it is specific to geometric analogy problems.

The Structure Mapping Engine (SME) \cite{falkenhainer1989structure} is an
analogy-making algorithm that gained notoriety for, among other things,
``discovering'' the Rutherford model of the atom by analogy to a solar
system~\cite{gentner1983structure}. Its basic operation is similar to that of a
sub-graph isomorphism algorithm, in that it looks for an injective mapping
between the objects and predicates in one structure into those of another that
retains the facts.

The SME algorithm has been criticized by some~\cite{hofstadter1995fluid} for
its reliance on hand-written input representations, which circumvents the
hardest part of analogy-making. For example, when forming an analogy between
the strings \stringobj{aaabbc} and \stringobj{abbccc}, it might be natural for
a human to group the three \stringobj{aaa}s and associate them either with the
single \stringobj{a} or the similar group-of-three \stringobj{ccc} in the
second. However, SME has no conception of modifying the structure, or its
representation of the structure, in such a way.  The user would have to
explicitly group letters before providing it to SME, at which
point SME would just look for relations between groups.

The \TC{} workspace represents a its workspace using structures and relations.
Similarly, TVLA~\cite{TVLA1,TVLA2,TVLA3} represents possible program states
using a variant of first-order logic. In TVLA, such structures are abstracted
into three-valued structures, somewhat similar to how we abstract concrete
instances in the workspace to form analogies.

\noindent\textbf{Comparative program understanding}
The problem of comparative program understanding is related to the problem of
code detection
\cite{roy2009comparison,marcus2001identification,kamiya2002ccfinder,DBLP:conf/sas/KomondoorH01,DBLP:conf/icse/GabelJS08}
with recent approaches using deep learning~\cite{white2016deep}. CP-Miner
identifies bugs related to copy-pasted code~\cite{li2006cp}. The func2vec
technique computes function embeddings to learn function synonyms, which are
functions are play a similar role in the source
code~\cite{DBLP:conf/sigsoft/DeFreezTR18}. Such function synonyms are used to
identify error-handling bugs in the Linux kernel. The code2vec technique
computes an embedding of source code using the
AST~\cite{DBLP:journals/pacmpl/AlonZLY19}, which can be used to infer names of
functions. These techniques represent a research thread that uses ``Big
Code''~\cite{DBLP:journals/ftpl/VechevY16}. In contrast, $\TC$ uses relatively
limited amount of source code to make analogies; however, it can make use of
models of source code learned via deep learning and other techniques.

\noindent\textbf{Program transformation-learning}
Recent approaches, such as GetAFix~\cite{bader2019getafix}, have explored the use
of \emph{antiunification} of program ASTs to learn program fixes. Such
antiunification can be seen as a restricted special case of the
analogy-as-abstraction process used by \TC{}
(see~\pref{sec:RepresentingAnalogies}). Existing antiunification-based
approaches, however, are limited to tree structures (such as ASTs), and can not
make use of additional semantic information. For example, GetAFix would not
have been able to accurately learn the transformation in \pref{sec:WhatIfGEMM},
which relied on semantic properties of the code, or the one in
\pref{sec:APIMigration}, which relied on referencing the documentation.

The program transformation-learning problem is related to Programming by
Example~\cite{DBLP:journals/pvldb/Singh16,DBLP:conf/popl/SinghG16,DBLP:conf/aaai/RazaG18,DBLP:conf/aplas/GulwaniJ17,DBLP:conf/ppdp/Gulwani17,DBLP:conf/popl/Gulwani11,DBLP:journals/pvldb/SinghG12,pygmalion,zloof1977query,miltner:blue_pencil}.
Such techniques typically restrict the program transformation to a limited
domain-specific language. Repenning et al.~\cite{repenning2000programming}
describe how end-users of a programming-by-example system might use analogies to
express the desired behavior by comparison to that of an existing program.
Perrone et al.~\cite{perrone1998graphical} propose that implementing code reuse
via concrete analogies can be more natural than the use of standard
object-oriented programming paradigms and help novice programmers avoid
copy-and-pasting code.
Recent approaches have explored using natural language and examples as input \cite{DBLP:conf/ijcai/RazaGM15}
and using deep learning techniques~\cite{DBLP:conf/iclr/VasicKMBS19}.

\noindent\textbf{API Migration}
Many approaches for API migration that use statistical and machine learning
techniques have been
proposed~\cite{DBLP:conf/icse/PhanNNN17,DBLP:conf/icse/NguyenNPN17,DBLP:conf/icse/NguyenNN16a,DBLP:conf/kbse/Nguyen16,DBLP:conf/icse/NguyenNNN14a,DBLP:conf/kbse/NguyenNNN14}.

\noindent\textbf{Cognitive Science in SE}
There are numerous works which have highlighted the promise of models from
cognitive science in software engineering. Call by
Meaning~\cite{callbymeaning} describes a system in which
program components (e.g., functions) can be addressed by their
\emph{semantic meaning,} not just their syntactic name. A programmer may
describe an existing function using a high-level, semantic description
language and the Cyc~\cite{cyc} cognitive model will be used to infer which
(composition of) function(s) best matches that description. The Semprola
semiotic programming language~\cite{semprola} allows programmers to
directly use \emph{signs} instead of the now-dominant focus on textual code
symbols. Both Call by Meaning and Semprola are ambitious and exciting
projects, requiring a fundamental re-thinking of how and in what languages
we write code. While we envision that \TC{} can benefit such approaches in
the future, we are excited that, as described in~\pref{sec:WhatIf}, \TC{}
can wield these cognitive models to more immediately benefit software
engineers using the existing programming languages and environments of
today.

\section{Conclusion}
In this paper, we discussed \emph{analogy-making}, a fundamental human ability
that involves identifying underlying similarities between two objects. We
first described analogy-making through examples in a restricted letter-string
domain. We then showed how analogy-making can be used to address a variety of
software engineering tasks, namely comparative program understanding, program
optimization, and API migration. Finally, we described \TC{}, our proposed
algorithm for analogy-making, which is suitable for making analogies about
programs. \TC{} relies on a novel triplet-structure representation for its
workspace, allowing it to form analogies over arbitrary inputs, such as source
code, program analyzer outputs, and documentation. By reducing a variety of
problems to analogy-making, improvements to the core analogy-making primitive
can pay large dividends across a variety of applications.  Software engineering
represents a difficult challenge for analogy-making, as it involves a unique
balance of unambiguous syntax and semantics of the program, as well as
ambiguous information about programmer intent.  We hope that this paper
serves as a first step towards cementing analogy-making as a core primitive in
the software engineering toolbox.

\begin{acks}
    We thank the anonymous reviewers and Cindy Rubio Gonz\'alez for comments
    that significantly improved this paper.
\end{acks}

\bibliographystyle{ACM-Reference-Format}
\bibliography{main}

\clearpage
\appendix
\section{Efficient Implementation of \TC{}}
\label{app:Implementation}
\subsection{Efficient Rule Matching in Triplet Structures}
Most operations on the workspace need to quickly look for patterns in the
triplet structure. Therefore, it is useful to represent the triplet structure
in such a way that many lookups are fast. Thankfully, because of the uniformity
of the triplet representation, we can do this. Internally, we represent triplet
structures by a hashmap, which takes \emph{triplets with holes} to a list of
triplets. In the example from~\pref{sec:ImplementationTripletStructures}, we
would associate with the key $(?, ?, \NextToLeft)$ the set of facts $\{ (f_1,
a, \NextToLeft), (f_2, b, \NextToLeft) \}$.  When adding a new fact to the
structure, we add it to the $2^3 = 8$ hashmaps formed by replacing some subset
of its indices with a hole.  This is a relatively manageable constant-factor
overhead.

With this representation, we can reduce looking for a node satisfying some
pattern to simply intersecting sets. For example, if we wanted to find a node
$v$ which is ``in the middle of a string,'' i.e.\ to the left of some node and
to the right of another, we would use the constraints $(?, v, \NextToLeft)$ and
$(?, v, \NextToRight)$. Since these are existential constraints in only a
single variable, we can solve them by intersecting
\[
    \begin{aligned}
        &\{ t_2 \mid (t_1, t_2, t_3) \in H[(?, ?, \NextToLeft)] \} \\
        \cap\quad
        &\{ t_2 \mid (t_1, t_2, t_3) \in H[(?, ?, \NextToRight)] \}.
    \end{aligned}
\]

Solving more complex constraints (e.g., finding multiple nodes which together
satisfy some constraints) is still an NP-hard constraint satisfaction problem
in triplet structures, but such efficient single-variable existential lookups
helps form the core of our solver for more complex constraint problems.

\subsection{Differential, Symmetric, and Local Rule Matching}
In addition to efficiently storing the triplet structures, \TC{} uses the following three
other optimizations to speed up rule pattern matching.

First, \TC{} makes use of \emph{differential matching.} The fundamental
observation is that, at least for the first existential layer of a rule, adding
facts to the structure can only ever add more possible rule assignments, so
every such new rule assignment must use one of the newly-added facts. Hence, \TC{}
can keep track of which facts have been added to the structure since it last
looked for rule assignments, and only consider assignments which use those new
facts. A similar technique works for filtering out old assignments that use
removed facts.

Second, we note that many rules have \emph{symmetries}, where a valid variable
assignment can be permuted to form a new valid variable assignment. Under
certain conditions, \TC{} can take advantage of these symmetries to only search
for assignments to half of the variables, then consider all permutations to get
the rest of the variables.

Finally, heuristics can be used to \emph{localize} the rules. For example, \TC{}
can only look for rules that match to symbols nearby a symbol modified
on the last update rule application (this is similar to the operation of the
Slipnet in Copycat).

\subsection{Commutative Node Names}
Many interesting search heuristics require reasoning across different branches
of the search tree. For example, \emph{phase saving} is used by SAT solvers
such as Chaff~\cite{moskewicz2001chaff}, where after backtracking and choosing
another assignment to variable $v_i$, the solver will attempt to perform the
same assignments to $v_j$ for $j > i$ as were made before the backtrack.

To apply analogous heuristics to the type of search over update rule
applications in \TC{}, we need to be particularly careful about how we
reference nodes in the structure. This is because rules can operate on nodes
created by other rules. If we give newly-created nodes random names, or use
something like a global counter to name them uniquely, then when trying to
re-apply any rules that referenced nodes created after the point of
backtracking will fail, because those exact nodes no longer exist, even if
nodes playing exactly the same role were created exactly the same way in the
new branch.

To address this issue, we name nodes according to a hash of the rule and
pattern assignment that produced them.  Modulo hash collisions and applying
the same rule to the same assignment twice, this makes node names commutative
with the exact order of applying update rules and allows us to attempt the
same rule assignment before and after a backtrack.

An alternative to this solution would be to explicitly keep information about
the provenance of such generated nodes and refer to it when necessary. If such
information is further stored in the structure itself, it could be used to
perform Metacat~\cite{marshall2000metacat} style meta-reasoning, as discussed
in~\pref{sec:HeuristicsForSWA}.

\end{document}